# Bacterial proliferation pattern formation

John S. Chuang,[1,*] Riccardo Rao,[1,2,3,4,*] and Stanislas Leibler[1,2]

[1] *The Rockefeller University, Laboratory of Living Matter, New York, NY 10065, U.S.A.*
[2] *Institute for Advanced Study, Simons Center for Systems Biology, School of Natural Sciences, Princeton, NJ 08540, U.S.A.*
[3] *Medical University of Vienna, Center for Medical Data Science, Institute of the Science of Complex Systems, Spitalgasse 23, 1090 Vienna, Austria*
[4] *Complexity Science Hub, Josefstädter Strasse 39, 1080 Vienna, Austria*

Bacteria can form a great variety of spatially heterogeneous cell density patterns, ranging from simple concentric rings to dynamical spiral waves appearing in growing colonies. These pattern formation phenomena are important as they reflect how cellular processes such as metabolism operate in heterogeneous chemical environments. In the laboratory, they can be studied in simplified set-ups, where spatial gradients of oxygen and nutrients are externally imposed, and cells are immobilized in a gel matrix. An intriguing example, observed in such set-ups over 80 years ago, is the sequential formation of narrow bands of high cell density, taking place even for a clonal population. However, key aspects of the dynamics of band formation remained obscure. Using time-lapse imaging of replicate transparent columns in simplified growth media, we first quantify the precision of the positioning and timing of band formation. We also show that the appearance and position of different bands can be modulated independently. This "modularity" is suggested by the observation that different bands differ in their gene expression, and it is reproduced by a theoretical model based on the existence of internal metabolic states and the induction of a pH gradient. Finally, we can also modify the observed pattern formation by introducing genetic modifications that impair selected metabolic pathways. In our opinion, the possibility of precise measurements and controls, together with the simplicity and richness of the "proliferation pattern formation" phenomenon, can make it a model system to study the response of cellular processes to heterogeneous environments.

## INTRODUCTION

From soil horizons to sea strata, microbes are continuously exposed to spatiotemporally heterogeneous chemical environments [1–3]. Elucidating how microbes collectively behave in these environments is important given the prominent role of microbial metabolism in driving biogeochemical cycles [4–6]. However, studying collective microbial behaviors is difficult in the wild, which motivates experimentation in simplified laboratory settings, where key features of natural environments such as gradients of chemicals can be monitored or controlled [7–9]. In these settings, microbial pattern formation phenomena occupy a privileged position, as collective behaviors manifest themselves in the form of macroscopic spatial structures. Understanding how these structures emerge from microscopic interactions is an intriguing conceptual challenge.

Physically, microbial patterns consist of uneven distributions of cell activity (e.g. gene expression) [10–12] or cell densities. Patterns based on uneven cell densities typically arise by either heterogeneous growth or anisotropic cell migration, or a combination of both. Microbial pattern formation resulting primarily from cell motility received widespread attention in the last few decades. While colonizing nutrient-rich patches, or simply responding to chemo-attractants, expanding bacterial colonies can self-organize into bulls-eye patterns [13, 14], fractal-like shapes [15], or even extended, symmetrical arrays of spots or stripes [16, 17]. Mixtures of motile and non-motile bacterial species may further lead to fluctuating density patterns as the consequence of hydrodynamic interactions [18], or even to flower-like patterns as the consequence of mechanical interactions [19]. The latter type of interactions has also been implicated in the formation of spiral-like patterns [20]. Recently, by connecting quorum sensing [21–23] and bacterial chemotaxis gene networks, synthetic biologists were able to make motility responsive to cell density, and in this way create novel synthetic cell patterning systems [24–26] (see also Refs. [27–29]).

In contrast to motility-based patterning phenomena, patterns originating solely from heterogeneous microbial growth [30] have received little attention. These patterns are nonetheless important as they better reflect how cellular processes such as metabolism (which mediates biochemical interactions and drives growth) operate in heterogeneous chemical environments [31]. A notable example of such "proliferation pattern formation" (PPF) phenomena can be observed in *gel-stabilized transient gradient systems* [32]. In these systems, a population of a single bacterial species is embedded in a column of transparent gel, and exposed to opposing gradients of exogenously-added nutrients and oxygen. "Transient"—as opposed to "stationary"—refers to the crucial fact that these gradients vary within timescales comparable to cellular proliferation. In fact, rather than replicating homogeneously as nutrients are transported locally by diffusion, cells proliferate preferentially in isolated bands visible with the naked eye (see e.g. the results presented in Fig. 1 and Mov. S1). The spatially heterogeneous distribution of these high cell density bands, as well as their sequential appearance, excludes conventional pattern formation mechanisms, such as Turing instabilities [33]. The fact that microbes are immobilized in a gel matrix rules out mechanisms based on cell motility [34, 35], collective swarming [36, 37], or cell diffusion [38]. This simple PPF phenomenon was discovered by J. W. Williams in the 1930's [39, 40], and his experimental set-up was later refined by Wimpenny *et al.* [32], who conducted many studies of microbial systems exposed to gradients of chemicals [41]. These authors also recognized that energy and amino acid metabolism play an important role in the pattern formation [42].

* These authors contributed equally to this work.

However, in these earlier studies, the dynamics of band formation were not examined in depth, and key questions remain unanswered: To what extent is band formation the outcome of contingent events, i.e. to what extent it is reproducible? Is the process underlying band formation the same for each band? How does the observed pattern formation depend on the chemical composition of the growth medium? How much is it influenced by specific metabolic genes? To answer these questions, we developed gel-stabilized transient gradient systems amenable to quantitative, reproducible studies, as well as ways to probe gene expression in individual bands. These advancements enabled us to reveal the multifaceted nature of PPF phenomena. First, the shape of the pattern is highly reproducible despite the fact that the timing of band appearance may be quite variable. Second, the appearance and position of different sets of bands can be modulated separately by acting on different variables. This central feature, which we refer to as "modularity", is highlighted by the fact that cells in different bands differ in their gene expression, and it is also recapitulated by a mathematical model of cell proliferation based on the existence of internal metabolic states and their different response to external pH levels. Third, consistent with such modularity, a great variety of band patterns can be obtained through manipulations that impact the nature of the external nutrients, the expression of metabolic genes, or even a quasi-stable colony-level phenotype.

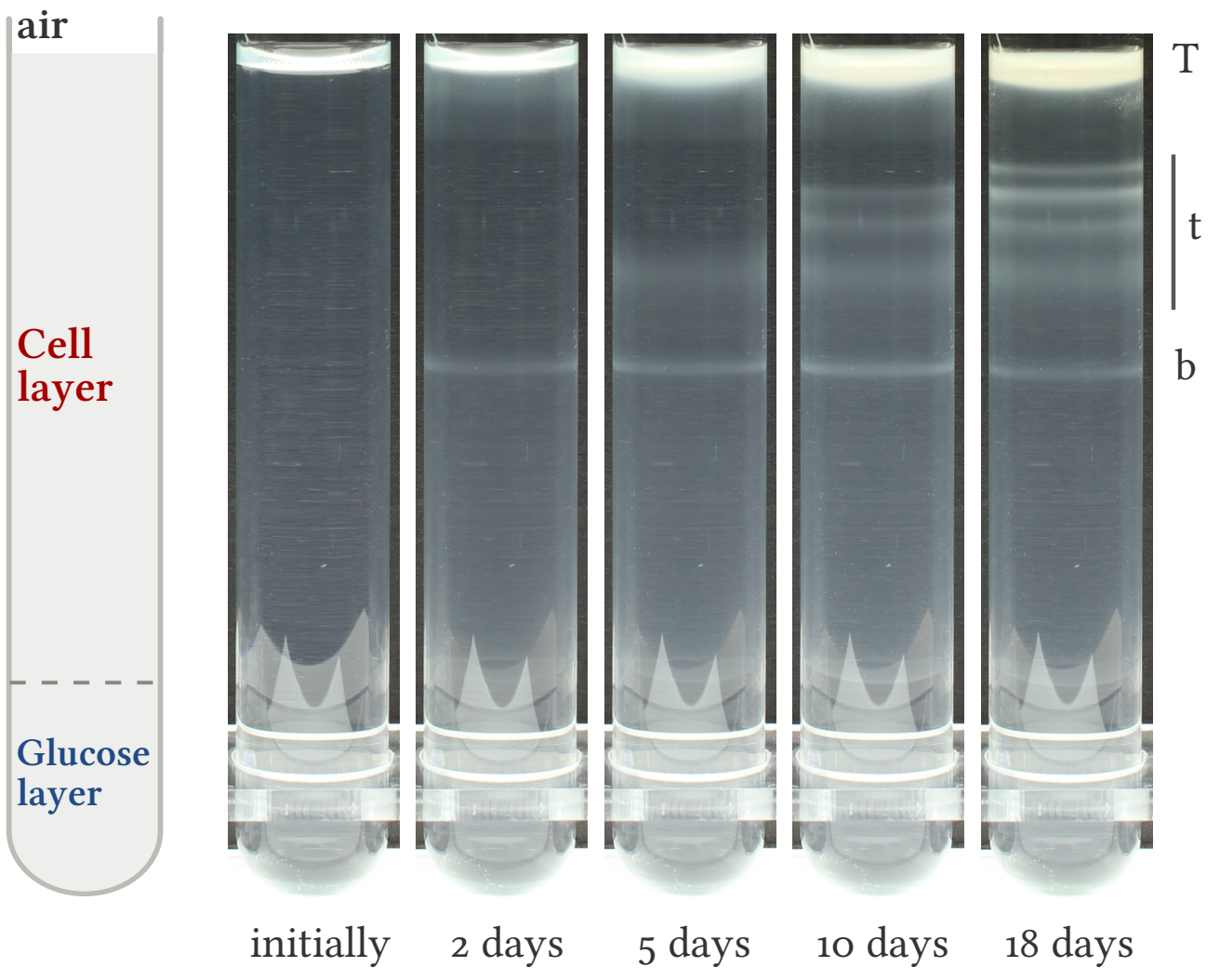

FIG. 1: **Schematic of a gel column and the PPF**. The two-layer gel system sits in a standard 16 × 150 mm test tube whose inner diameter is 13.55 ± 0.15 mm). A so-called bottom interface (dashed line) separates the Glc-layer (Glucose layer, in blue) from the C-layer (Cell layer, in red), whereas the top interface ("T") separates the C-layer from the air. The height of the typical C-layer is 69.5 ± 1.9 mm [43, §IB]. In this experiment, nitrogen is supplied to the medium in the form of a single amino acid, glutamate (additional details are discussed in [43, Fig. S1]; snapshots are extracted from Mov. S1). Among the bands, we highlight the bottom one ("b") from the top ones ("t").

## I. PPF IN GEL-STABILIZED TRANSIENT GRADIENT SYSTEMS

The pattern formation dynamics take place in a two-layer gel system with a total height of approx. 10 cm (Fig. 1, schematic column): a glucose-containing gel layer at the bottom, devoid of bacterial cells (*Glc-layer*); and another gel layer atop, initially containing no glucose, but inoculated with bacterial cells (*C-layer*). A growth medium containing mineral nutrients and a nitrogen source is initially distributed homogeneously in both layers (see Methods, App. A). Upon onset of the experiment, diffusion transports glucose from the Glc-layer to the C-layer above, enabling bacteria to grow on this energy-rich carbon source. Motility is impeded by the high viscosity of the gel, and indeed knocking out a flagella assembly gene has no effect on the pattern formation [43, Fig. S10].

The first band appears roughly halfway through the C-layer after 1–2 days, followed by a few additional bands appearing sequentially above the first. The PPF dynamics fully unfold within 10–20 days, where the precise duration depends on the medium composition, the initial amount of glucose, and importantly, on the conditions imposed at the top interface. Here, the C-layer can be either exposed to air or instead covered with a thick layer of mineral oil. In the former condition, which is the case of the experiment depicted in Fig. 1, top interface cells have access to oxygen, and can proliferate to high densities by aerobically catabolizing amino acids, i.e. using amino acids as an energy and carbon source in an oxygen-containing environment (see top of the columns in Fig. 1, highlighted with "T") [44].

Although PPF dynamics are not restricted to a specific bacterial species, the ability to grow anaerobically by fermentation of the diffusing energy source—here glucose—seems essential [32]. Evidence of the significance of fermentation for PPF is manifold. Here, it suffices to note that close to the interface between the two gel layers—which we call the *bottom interface*—the pH falls below approx. 5 within 12–48 hours (SI Mov. S2)—a time scale consistent with that of band appearance. If fermentation is impaired by knocking out relevant genes (e.g. *pflB*, [43, Fig. S11]), cells display no proliferation in the bulk of the column (although they do grow at the top interface), and no band formation is observed.

Unless stated otherwise, we report on the proliferation patterns realized by the facultative anaerobe *Serratia marcescens* (strain ATCC 13880).

## II. REPRODUCIBILITY

Assessing the reproducibility of PPF required making several technical improvements compared to previous experimen-tal set-ups, *cf.* [32]. We adopted a clearer gel (0.3% Phytagel), minimized evaporation (and thus gel shrinkage), and captured images containing 12 replicate columns using time-lapse photography. This enabled us to assess the spatiotemporal variations of the intensity of light scattered through each column—a proxy of cell density—with greater precision and resolution. We also devised a minimal synthetic medium with a defined composition consisting of amino acids (as the nitrogen source) and mineral nutrients. To our surprise, patterns form even in media containing a single amino acid such as glutamate or glutamine (see §IV and [43, Figs. S12–13]).

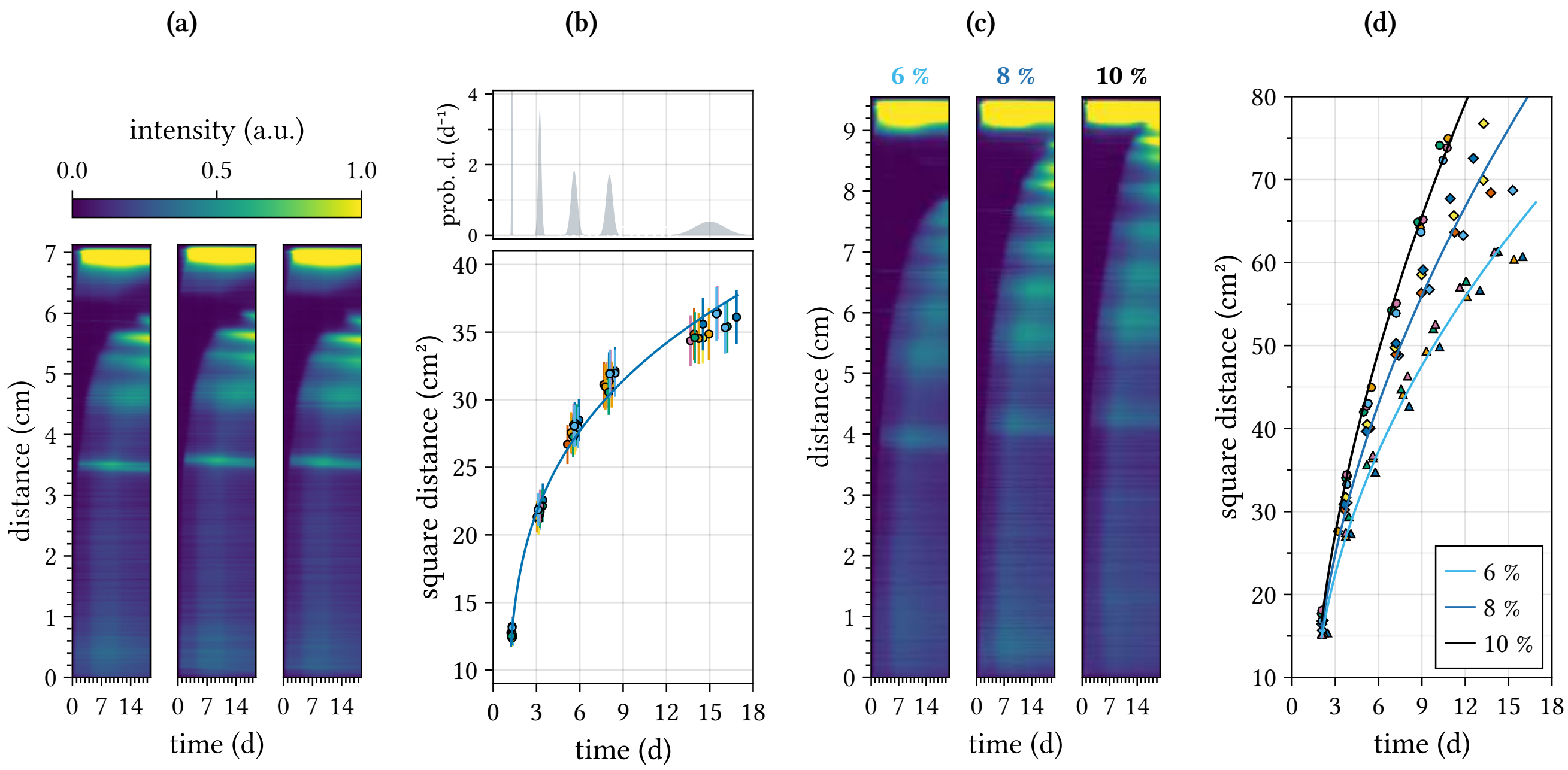

FIG. 2: **Reproducibility, variations, and spatiotemporal scaling of PPF. (a)** Kymographs of replicate PPF. On the vertical axis, "distance" refers to the distance from the bottom interface. The intensity profile is rescaled so that the maximum intensity detected in the bulk of the columns (region excluding the bottom and top interfaces) belonging to the same experiment (always presented as a separate sub-figure) is 1. This means that band intensities should be compared solely within a sub-figure. Kymographs in (a) are taken from a 12-replicate experiment [43, Fig. S1]. In **(b)**, we plot the spatiotemporal scaling of coordinates of band appearance for such experiment: band position squared vs band appearance time. The error bars represent the measurement uncertainty, whereas the blue solid line is a fit based on the model of anomalous diffusion introduced in the main text. Parameters are obtained from maximum likelihood (ML) estimation: $\alpha = 0.27 \pm 0.01$, $K = 18.1 \pm 0.3\,\mathrm{cm}^2\mathrm{d}^{-\alpha}$, $\tau = 1.03 \pm 0.03$ d ($r^2 \simeq 0.99$, [43, §III]). The top panel depicts Gaussian distributions with means and standard deviations equal to those of band appearance times (grouped by band ordinality). **(c)** Kymographs of PPF obtained when increasing the initial concentration of glucose in the Glc-layer, $[\mathrm{Glc}]_0 = 6, 8$, and 10 % (w/v) (1 % (w/v) = 0.01 g (solute) / mL (solution)). The 3 kymographs in (c) are taken from a 3×4-replicate experiment [43, Fig. S3]). In **(d)**, we plot band positions squared against appearance times: △, 6 %; ◇, 8 %; and ○ 10 % (w/v). The data points are modeled using the relation of anomalous diffusion introduced in the main text, but with $\alpha$ and $K$ changing linearly with the concentration of glucose [43, §III, Eq. (S3)]. ML estimation of the parameters (solid lines, $r^2 \simeq 0.99$) gives $\mathrm{d}\alpha/\mathrm{d}[\mathrm{Glc}]_0 = 0.020 \pm 0.001$ [% (w/v)]$^{-1}$.

At fixed medium composition, band patterns are basically indistinguishable (Fig. 2 and [43, §II]). The kymographs in Fig. 2a show the spatiotemporal dynamics of band formation in three representative columns from a 12-replicate experiment (additional details in caption, and Extended Methods [43, §IA], and [43, Fig. S1–2]). These graphical representations exploit the fact that the pattern formation is effectively one dimensional: the distance of a generic point from the bottom interface changes along the vertical axis, while time from the start of experiments increases along the horizontal axis. The intensity of the heat map quantifies the intensity of scattered light. The snapshots in Fig. 1 correspond to specific times (vertical slices) of the first kymograph in Fig. 2a.

Figure 2b (bottom panel) shows the square of band positions (distances from the bottom interface) vs the time of their appearance. Measurement uncertainties (the error bars) cover the statistical deviations of band positions. For the bottom band, this uncertainty is 0.9 mm, which corresponds to roughly $10^2$–$10^3$ *S. marcescens* cell lengths. Notwithstanding such high spatial reproducibility, band appearance times exhibit significant variations. Measured by standard deviations, these variations increase from less than 1 hour (for the bottom band) to more than 1 day (for the top-most band) (Fig. 2b, top panel, and [43, Figs. S1d and S2d]). We find this result surprising as it suggests that the spatial and temporal coordinates at which bands appear may not be tightly coupled.

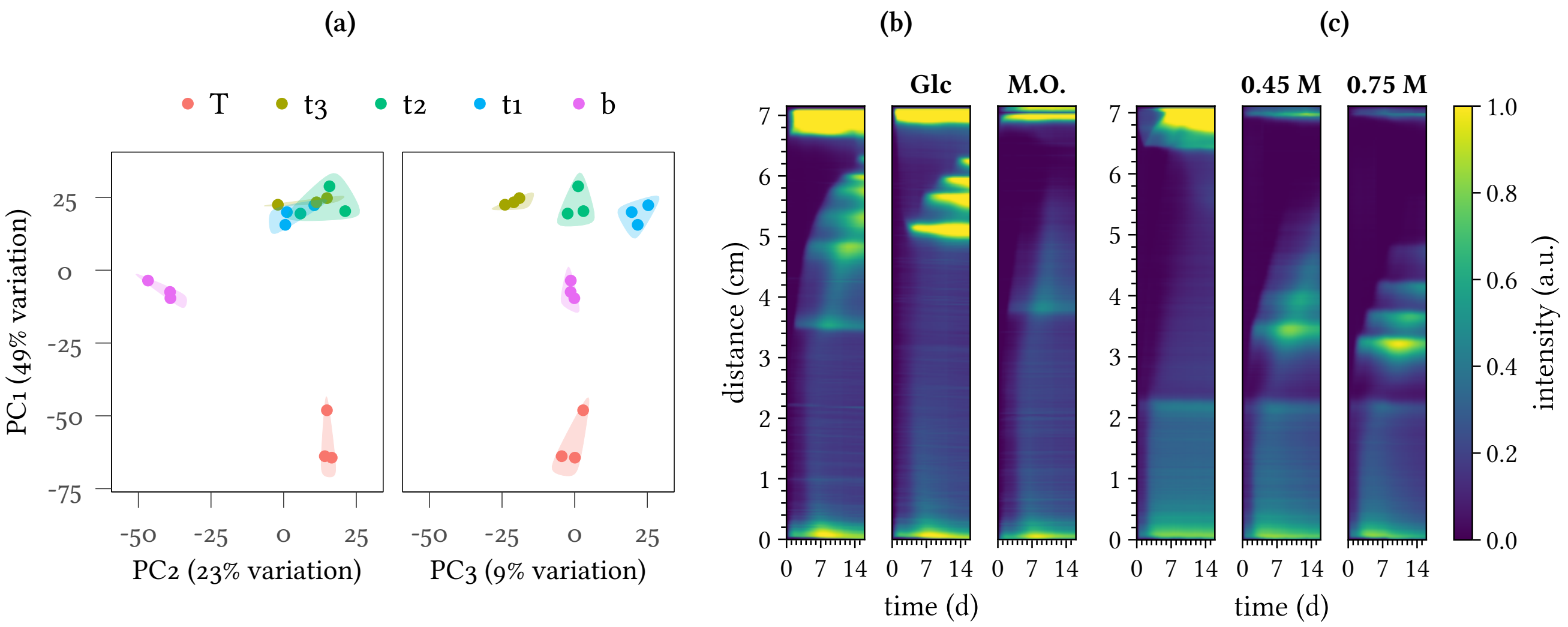

FIG. 3: **Modularity of PPF.** **(a)** Gene expression profiles of cells proliferating in bottom band (b), top bands (t1, t2, and t3; in sequential order of appearance), and at the top interface (T) projected on principal components 2 vs 1, and 3 vs 1. Shaded areas are visual aids grouping together the 3 replicate samples. Details about the experimental set up and procedure used to assess gene expression profiles are found in App. A and [43, §§IB and C]. **(b)** Kymographs comparing a control PPF and that obtained by *(i)* adding 0.05 % (w/v) glucose to the medium in the C-layer (Glc), and *(ii)* overlaying the C-layer with mineral oil (M.O.). For visualization purposes, the intensity is rescaled here so that the maximum value of the bands in the control (1st column) is set to 1. **(c)** Kymographs of PPF obtained by using ammonium ions rather than amino acids as nitrogen source ($NH_4Cl$). The 2nd and 3rd columns report the outcome of overlaying the C-layer with mineral oil and adding a buffer to the top interface (0.5 mL of 0.45 M or 0.75 M potassium phosphate buffer, pH 7, details in [43, Fig. S5]).

**Remark** Figure 2b also highlights how band coordinates align along a *scaling curve*, as if variations across different replicates were constrained. As band formation is likely triggered by glucose—which is transported by diffusion—the fact that these coordinates follow a scaling relation should not come as a surprise. However, the scaling is incompatible with canonical diffusion (Fig. 2b, blue line, and [43, §III]): positions, $x$, and times, $t$, scale as $x^2 = K(t-\tau)^\alpha$ with $\alpha$ statistically incompatible with the value 1, suggesting rather a sub-diffusive scaling. Here, $K$ is an anomalous diffusion coefficient and $\tau$ a time lag.

Arguably, one factor that contributes to make the scaling incompatible with canonical diffusion is glucose consumption by bacterial growth, which reduces the diffusion flux and delays glucose transport up the column. In fact, we observe that when the initial concentration of glucose is increased, the band scaling becomes closer to that of diffusion, viz. $\alpha$ increases (Fig. 2d and [43, §III]). In addition, bands appear later and higher up, which suggests that: *(i)* glucose sets the tempo of band appearance by triggering cell proliferation; and *(ii)* temporary and local glucose depletion is likely involved in the mechanism leading to band formation.

## III. PATTERN MODULARITY

A noticeable feature of the patterns shown in Figs. 1 and 2 is that the bottom band appears more distant from the second one than the top bands are among themselves. This observation is not restricted to specific conditions and questions the assumption that the mechanism leading to band formation is the same for all bands.

Seeking clarity, we investigated the gene expression profiles (transcriptome) of cells in newly-formed bands through RNA-seq (App. A and [43, §IB and C]). For completeness, we also profiled cells growing at the top interface. The result of 3 replicates each for the first 4 bands (b, t1, t2, and t3) and for the top interface (T), is summarized by the principal components analysis (PCA) plot in Fig. 3a, which shows the transcriptome projected on the first 3 principal components (PC). Notably, the first 2 PCs (capturing approx. 70% of the variance) discriminate not only the top interface from the bands, but also the bottom band from the top bands. This clearly shows that cells in different bands operate different genetic programs.

The above result motivated a search for phenomenological conditions in which the bottom band and top bands appear in isolation. We found that modifying the medium in the C-layer through the addition of small amounts of glucose disrupts the formation of the bottom band but not the remaining bands (Fig. 3b, 2nd kymograph). In contrast, overlaying the top layer with mineral oil, and thus limiting the aerobic catabolism of amino acids at the top interface, disrupts the top bands but preserves a dim bottom band (Fig. 3b, 3rd kymograph).

To explain this behavior, we recall that proliferation of the bacterial cells at the top interface hinges upon using amino acids not only as a source of nitrogen, but also as a source of energy and carbon. However, this proliferation regime requires less nitrogen than what amino acids provide, and the excess is released in the form of ammonia ($NH_3$) [44]. Diffusing down the column, $NH_3$ forms a spatial gradient that is likely essential for the formation of the top bands. In principle this may happen for many reasons [43, §IV], but we argue that one is dominant: being a weak base, $NH_3$ counteracts the acids released by glucose fermentation, and thus neutralizes pH. Arguments supporting this hypothesis are manifold [43, §IV], but we report here solely the most compelling one: replacing the amino acids in the medium with ammonium ions (specifically $NH_4Cl$, offering no buffering effect) disrupts the formation of the top bands (Fig. 3c, 1st kymograph). Yet, top bands can be re-established upon exogenous addition of a (nitrogen-free) buffer at the top interface (Fig. 3c, 2nd and 3rd kymographs).

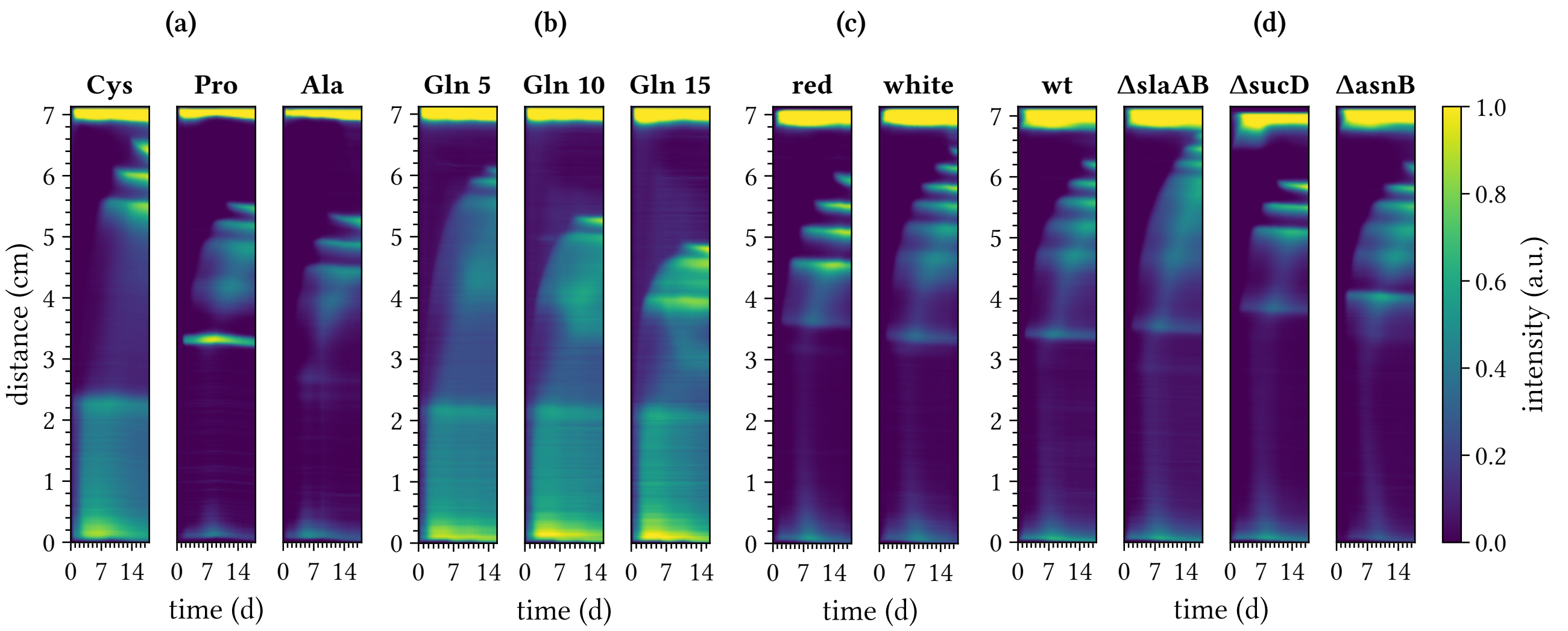

FIG. 4: **Pattern manipulations (a)** Kymographs of 3 PPF in which the growth medium is prepared with 3 different single amino acids: Cys, cysteine; Pro, proline; and Ala, alanine. Additional amino acids and details are presented in [43, Fig. S12]. **(b)** Kymographs of 3 PPF in which the growth medium is prepared with glutamine (Gln), but in different initial concentrations: 5, 10, and 15 mM. Additional glutamine concentrations and details are presented in [43, Fig. S4]. **(c)** Kymographs of 2 PPF in which the C-layer is inoculated with 2 distinct phenotypes sampled from a clonal population: red, prodigiosin-production competent; white, prodigiosin-production incompetent. Additional replicates and details are presented in [43, Fig. S14]. **(d)** Kymographs of 4 PPF in which the C-layer is inoculated with the wild type strain (wt), *ΔslaAB*, *ΔsucD*, or *ΔasnB* knockout mutants. Additional mutants, replicates, and details are presented in [43, Fig. S15].

Overall, these results indicate that countergradients of acids—or protons—and a base (either endogenously produced or exogenously provided) are essential for the formation of the top bands, but not necessarily so for the bottom one. This conclusion suggests the possibility of modulating different components of proliferation patterns by manipulating the chemical medium. In fact, we do observe significantly different patterns when preparing the medium with different single amino acids (Fig. 4a, and [43, Fig. S12]). Increasing the initial amino acid concentration shifts the top bands downward in a progressive fashion (Fig. 4b, and [43, Fig. S13]), which is likely the outcome of steeper gradients of ammonia. This observation is consistent with what is observed in Fig. 3c (2nd and 3rd kymographs) where the buffer, rather than ammonia, is increased.

**Remark** The experiment presented in Fig. 3c also implicates a second factor contributing to make the scaling of PPF incompatible with canonical diffusion: the countergradient of base. We indeed find that increasing the concentration of buffer decreases the scaling exponent $\alpha$ [43, §IV and Fig. S5b].

## IV. NON-GENETIC AND GENETIC MANIPULATIONS OF PPF

We next sought to ascertain to what extent PPF can be manipulated via endogenous manipulations of cellular metabolism, rather than exogenous ones. Two types of changes can endogenously impact bacterial metabolism: those connected with phenotypic switching and those affecting metabolic genes, e.g. mutations or deletions.

Clonal populations of *S. marcescens* ATCC 13880 can produce 2 quasi-stable colony-color phenotypes (red vs white) differing by their ability to synthesize the red pigment prodigiosin, a secondary metabolite that incorporates several amino acids as precursors [45]. This is a classic example of epigenetic (*sensu* C. H. Waddington) phenotypic switching [46]. To our surprise, gel columns inoculated with cells from red versus white colonies displayed consistently distinguishable patterns (Figs. 4c and S14).

Similarly, we found that knocking out genes involved in anaerobic fermentation, aerobic catabolism, or nitrogen metabolism generally alters the pattern (Figs. 4d and S15).

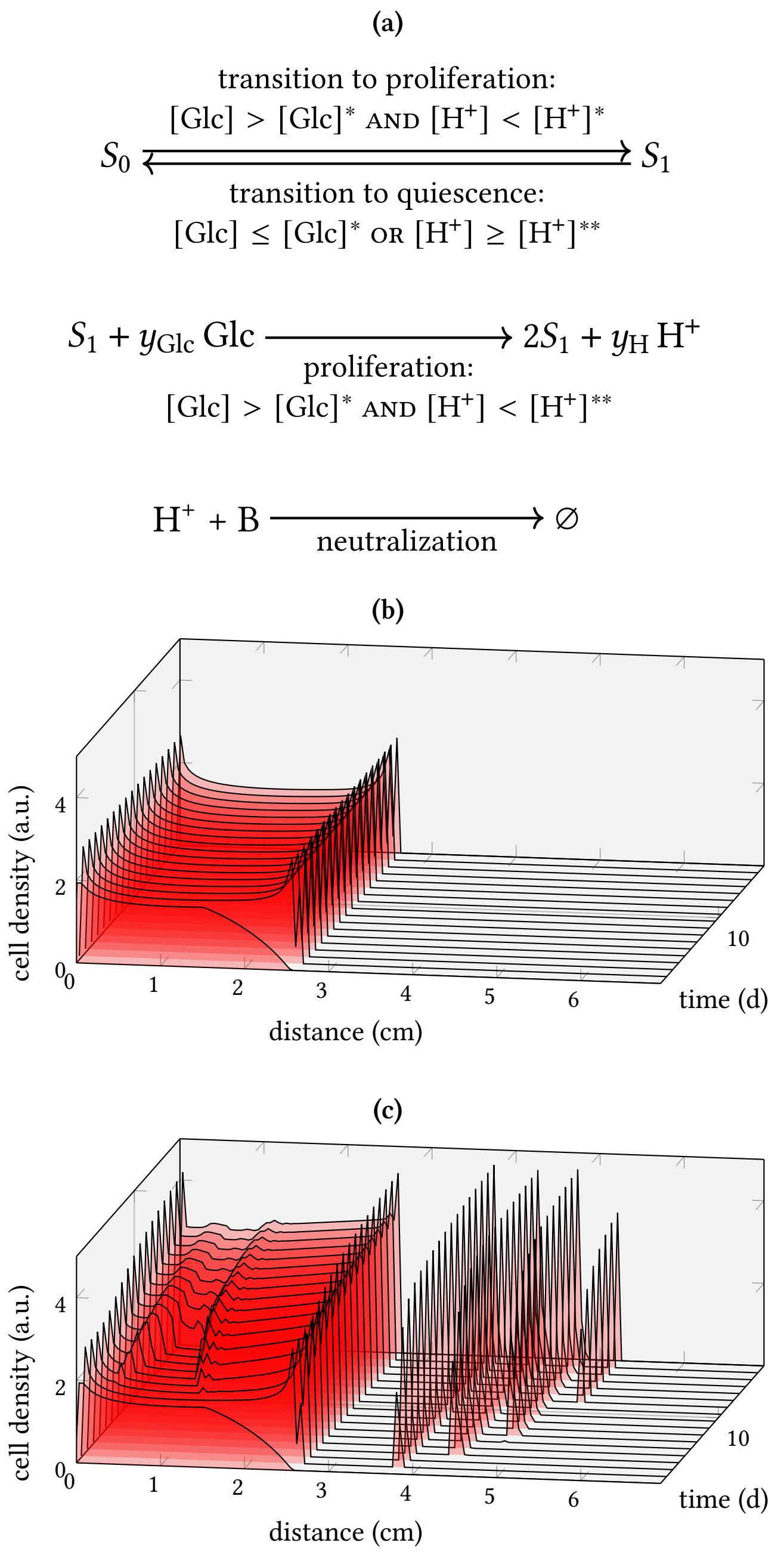

FIG. 5: **Model schematic and results (a)** Essential schematic of the model: $S_0$ and $S_1$ represent quiescent and proliferating cell states, respectively; Glc, $H^+$, and B, represent glucose, protons, and the base, respectively. Cell state transitions and proliferation are active/inactive *conditionally* on the availability of glucose and the concentration of protons, i.e. pH. $[\mathrm{Glc}]^*$, $[\mathrm{H}^+]^*$, and $[\mathrm{H}^+]^{**}$ denote the threshold concentrations. Proliferation consumes $y_{\mathrm{Glc}}$ amount of glucose and produces $y_{\mathrm{H}}$ amount of protons. The last process describes the reaction of neutralization, with ∅ being an inert, or "null", chemical. Glc, $H^+$, and B are subject to diffusion, whereas cells—no matter the state—are not. **(b)** and **(c)** Simulations of the PDE-based model built on the schematics in (a) either without (b) or with (c) an exogenously-added base B diffusing from the top interface.

We illustrate these three cases via *ΔslaAB*, *ΔsucD*, and *ΔasnB* mutants, respectively. The *slaAB* genes are essential for operating the 2,3-butanediol fermentation pathway, i.e. a pathway present in some bacteria such as *S. marcescens.* Knocking out *slaAB* forces fermenting cells to release stronger acids [47], and the bands appear less sharp and are shifted upwards (Fig. 4d, 2nd kymograph). Knocking out *sucD*, which encodes the TCA cycle enzyme succinyl-CoA synthetase, active during aerobic respiration, results in delayed proliferation of top interface cells, and although sharper, bands appear shifted up the column (Fig. 4d, 3rd kymograph) relative to wild-type. Finally, *asnB* is a gene whose product is involved in asparagine synthesis and ammonia assimilation pathways. In this case, we observe that mostly the bottom band is affected, appearing shifted upwards (Fig. 4d, 4th kymograph)—yet another illustration of how bottom band and top bands are differentially modulated.

## V. MODEL OF PPF

Having highlighted the role of glucose, pH, and a base countergradient we next aimed at clarifying the mechanism behind the formation of proliferation patterns, as well as their modularity. Simplistic models based on exponential or logistic growth happening when both glucose and pH lie above certain thresholds fail to reproduce the band formation. We argue that one key aspect is neglected in those models: cells transition between different physiological states upon changes of environmental conditions, and respond differently depending on which state they are.

The following model (schematized in Fig. 5a, and formulated mathematically in App. B) incorporates this ingredient and qualitatively reproduces PPF phenomena. First, cells are initially in a quiescent state ($S_0$) and then transition to a proliferating state ($S_1$) upon arrival of glucose (Glc) in a small but finite amount. Such transition happens provided that the proton concentration $[\mathrm{H}^+]$ is below a certain threshold, denoted by $[\mathrm{H}^+]^*$ (viz. the external medium is not too acidic). Second, while proliferating, cells release protons as consequence of their fermentation. When the proton concentration exceeds a higher, second threshold, $[\mathrm{H}^+]^{**}$, proliferation halts. As a consequence of the assumption $[\mathrm{H}^+]^{**} > [\mathrm{H}^+]^*$, there exists a range of $[\mathrm{H}^+]$ where replicating cells keep replicating while quiescent cells remain quiescent. Finally, when conditions are not suitable for proliferation (because of either lack of glucose, or too acidic an environment), cells revert to the quiescent state, $S_0$.

With parameters coarsely adjusted within biologically-plausible ranges [43, Tab. S6], and initial and boundary conditions reproducing those of the experiment in Fig. 3c (App. B), our model captures not only band formation, but also the independent formation of the bottom band and top bands (Figs. 5b and 5c; see [43, Fig. S16] for a qualitative comparison with experiments). In addition, our model qualitatively recapitulates how patterns respond to perturbations that can be studied experimentally, such as changes of the initial concentrations of glucose, or even the amount of protons released by cells upon proliferation [43, §VA]).

## DISCUSSION

### Reproducibility of pattern formation

Improving upon the gel-stabilized transient gradient system [32], we have shown that the formation of proliferation patterns is generally robust to drastic simplifications of the growth medium, such as using single amino acids as the nitrogen source (Figs. 4a and S12). The simplification and control of the media composition also enabled us to show that replicate patterns are remarkably alike (Fig. 2a and [43, §II]). In contrast to low spatial variations of band appearance, temporal variations may be significant, especially for the top-most bands. This implies thae: *(i)* the variables that control band positions might be different from those that control band appearance times, i.e. cellular proliferation is spatially constrained by some "morphogens" (*sensu* A. M. Turing [33, 48]) but temporally might be activated by others. *(ii)* these latter morphogens, or the cellular programs activated by them, may be subject to variations that increase as the pattern formation unfolds. One possible scenario is that the countergradients of acids and bases—generated respectively by anaerobic glucose fermentation and oxidative amino acid catabolism—control the position of band formation, whereas the arrival of glucose transported by diffusion controls the cell proliferation leading to band appearance. Contingent variations of glucose arrival times (possibly connected to differences of glucose consumption rates or small differences of initial glucose concentrations), or of the activation of cellular proliferation, would then lead to temporal variations of band formation. Unfortunately, testing this scenario would require significant modification of our experimental set-up, and hence lies beyond the scope of this paper.

### Local activation, lateral inhibition

A somewhat surprising outcome of our investigation is the possibility of modulating the appearance and position of different bands by altering the growth medium or the conditions imposed at the top interface (Figs. 3 and 4). This result is reinforced by the observation that cells in different bands operate different cellular programs (i.e. exhibit different gene expressions), and seems to suggest that different bands form via different mechanisms.

Our model offers a more general explanation: all bands form as the outcome of the same type of process, whose principle is broadly known as *local activation–lateral inhibition* [49]. Cell proliferation is locally activated upon arrival of glucose, which is transported by diffusion. Being exponential, proliferation consumes glucose faster than the rate at which diffusion can replenish it. Hence, the glucose diffusion flux quickly decreases, and the glucose-induced transition to proliferation cannot take place further up the column (downstream with respect to glucose diffusion). At the same time, proliferating cells also release protons, which add to those released by cells below (closer to the bottom interface) and eventually lead to a halt of proliferation and its activation. These two effects—glucose depletion up the column and high proton concentration down the column—laterally inhibit proliferation or its activation and lead to band formation. Without any pH-neutralizing agent, protons are released in too high an amount to enable the formation of more than one band. The presence of a source of diffusing base at the top interface thus enables top band formation at the moving countergradient interface.

Wimpenny *et al.* recognized early the main mechanism of band formation and built a theoretical model to illustrate it [50, 51]. Their model was inspired by existing models of the so-called Liesegang patterns [52, 53]. In Liesegang patterning, sequentially appearing bands of precipitating salt form when one electrolyte diffuses from one boundary and the other is initially homogeneously distributed. The analogy between PPF and Liesegang patterns is indeed striking, although phenomenological aspects such as the scaling of band appearance coordinates—subdiffusive for the former (Fig. 2b), diffusive for the latter [52, 53]—distinguish them. Independently inspired by Liesegang phenomena, our model also builds on the idea of multiple cell states and their differential response to morphogen concentration. However, our model considers protons as a morphogen responsible for the lateral inhibition, rather than just an essential resource that is depleted upon cell proliferation. This enables us to explain the observed modularity—a feature previously unappreciated—and find closer agreement with experiments.

A seemingly important feature of protons is that their diffusivity is much larger than that of glucose—the "activator" morphogen. This situation is reminiscent of the condition required for Turing patterning [33]. One may ask here whether having morphogens with substantially different diffusivities is not only important for a qualitative description of the proliferation pattern formation phenomenon, but also essential for the band formation itself. We find this not to be the case: bands can form even with diffusion coefficients of comparable magnitudes. We also find that, depending on whether the condition $D_{\mathrm{H}}\, y_{\mathrm{H}}^{-1} \gg D_{\mathrm{Glc}}$ is satisfied or not, bands typically do or do not form, respectively ($D$ and $y$ denote diffusion and stoichiometric coefficients, respectively, see Fig. 5a; further details in [43, §VB]). A precise characterization of the conditions for band formation is an interesting problem which we leave for a future study.

Finally, our model also clarifies the distinction between morphogens that are essential for the pattern formation—such as glucose and protons—from those that modulate it. Here, the role of "modulator" [54] is played by the diffusing base.

## PERSPECTIVES

The pattern modularity (Fig. 3) and the wealth of shapes that ensues (Fig. 4) suggest a novel paradigm of "programming pattern formation" [27, 28]: a specific pattern can be in large extent predetermined by a suitable choice of the medium composition and of the metabolic competences of bacterial cells, i.e. genetic programs of their metabolism. If this choice cannot be fully determined in advance, it could be possible to search for it *in silico*, upon development of quantitative models of PPF. One could also contemplate the idea of engendering specific patterns through *in silico* evolution, as is currently done for proteins [55], i.e. by generating random variations in the parameters associated with the composition of the medium or cellular metabolism, and then selecting patterns closer to the desired one. Beyond these ideas, we see the pursuit of programming specific PPF as a path towards a more fundamental understanding of bacterial physiological responses to heterogeneous environments.

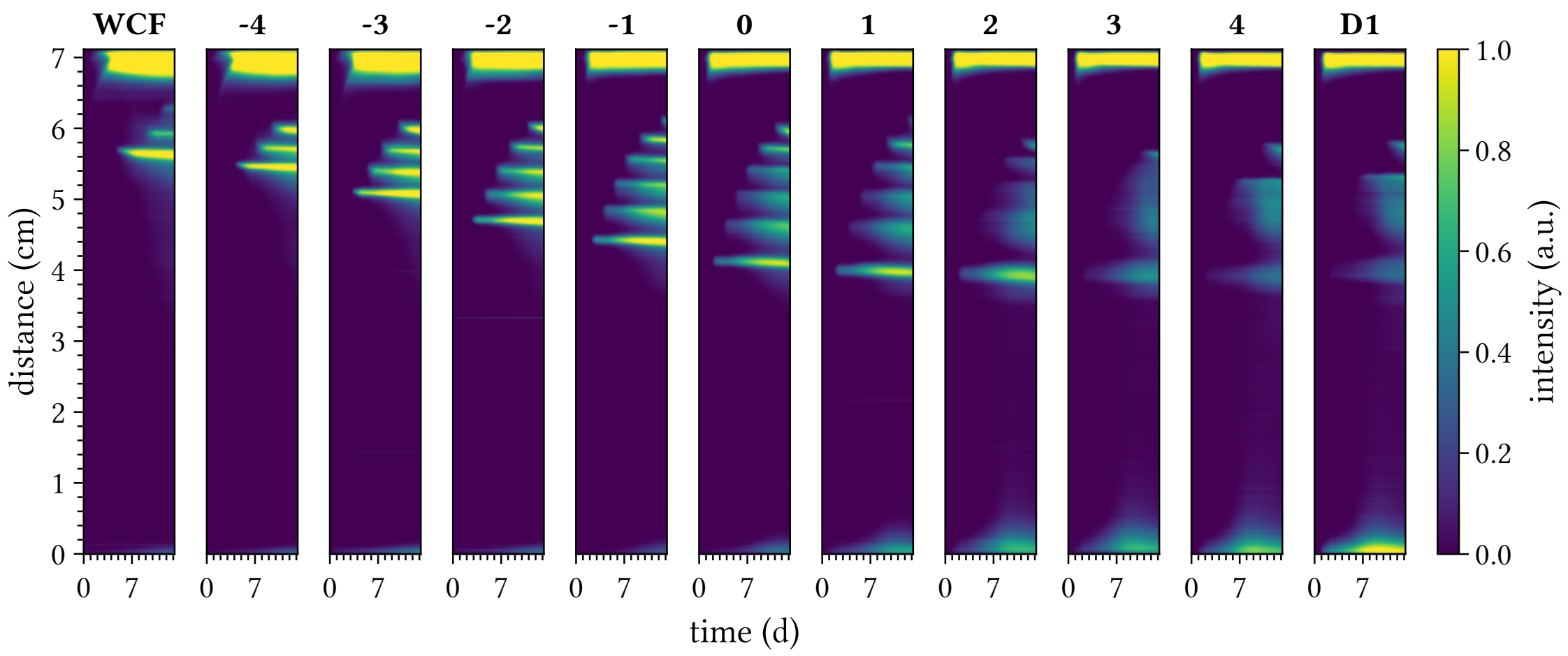

FIG. 6: **Manipulating PPF by mixing 2 strains in varying proportions** The first and last kymographs show the patterns realized by *S. marcescens* strains WCF and D1, respectively (details in App. A and [43, Tab. S1]). Each intermediate kymograph is the result of mixing these strains while keeping the overall initial density constant. The numeric labels on top represent the base 10 logarithm of the ratio of the proportion of D1 over WCF, $\log p_{\mathrm{D1}} - \log p_{\mathrm{WCF}}$. For visualization purposes, the intensity is rescaled here so that the maximum value of the bands in the evenly mixed column (6th kymograph) is set to 1.

Generating specific proliferation patterns may not be limited to manipulating the medium or the metabolic capabilities of the bacterial strain. We also find that mixing two different strains of *S. marcescens* in varying proportions—while keeping the initial overall cell density constant—leads to a progressive transformation of the pattern (Fig. 6). This study could be extended to explore whether there might be any compositional rules that govern the proliferation pattern for mixtures of different strains or species.

Within the realm of multi-species systems, it might be worth addressing the question whether one could build minimal persistent self-organizing (patterning) systems analogous to closed microbial ecosystems studied in Refs. [56–58] for motile organisms. Pursuing these endeavors may facilitate bridging the gap between gradient systems and so-called Winogradsky columns [59, 60], *viz.* complex motile microbial communities which self-stratify driven by endogenously generated gradients of inorganic chemicals and metabolites. It is important to note, however (as highlighted by the experiment presented in Fig. 6), that moving to multiple-species systems involves possibly non-trivial technical challenges. Moreover, it raises important theoretical challenges, such as finding suitable low-dimensional descriptions [56, 61].

## CONCLUSIONS

In summary, PPF dynamics contrast with many studies of microbial pattern formation phenomena, because no cell motility is involved here. Crucially, primary, rather than secondary, metabolism is the key player. Bringing this feature into play, we were able to modulate proliferation patterns by either manipulating the chemical media, or genetically modifying the metabolism of the bacterial strain, or even combining different strains. These results hint at the possibility of programming microbial patterns, and more importantly indicate that PPF can be adopted as a model system to study the interplay between internal cellular processes (i.e. genetic, epigenetic, metabolic, ...) and external spatio-temporally heterogeneous chemical environments.

## ACKNOWLEDGMENTS

We thank D. Forastiere for insightful feedback on the manuscript. This work has been partly supported by grants from the Simons Foundation (to S.L.) through the Rockefeller University (grant 345430) and the Institute for Advanced Study (grant 345801), and by grants from the Simons Foundation (#691552, to R.R.), and the Austrian Science Fund FWF (P34994, to R.R.).

## Appendix A: Methods

**Strains and Chemicals** Phytagel (aka gellan gum) was from Sigma (P8169). L-glutamic acid, monopotassium salt monohydrate (aka glutamate) was from Sigma (G1501). Other chemicals were purchased from Sigma, Fisher, or VWR. The *Serratia marcescens* type strain ATCC 13880 (Microbiologics KWIK-STIK format) was purchased from VWR. Strains D1 and WCF were from Carolina Biological Supply. Mutant strains of *Serratia marcescens* were constructed using the pTOX allelic exchange vector system [62], as described in detail in the SI Methods [43]. See [43, Tab. S1] for a listing of strains used.

**Phytagel Column Preparation** To pour Phytagel columns for bacterial growth, filter-sterilized media components were added one at a time, with swirling after each addition, into previously autoclaved solutions of Phytagel in water pre-warmed either to 50°C (Glc-layer, bottom) or 42°C (C-layer, top), with D-glucose present only in the Glc-layer (see below). The final concentrations were 0.3% (w/v) Phytagel, 4.4 mM $K_2HPO_4$, 0.2 mM $MgSO_4$, 25 µM $MnCl_2$, 7 µM $CaCl_2$, 7 µM $FeSO_4$, 6.8 mM glutamate, with an additional 4 mM $MgSO_4$ added last while swirling the prewarmed media to initiate gelation. In some columns, we also added 5 µM $Na_2MoO_4$, 5 µM $H_2SeO_3$, 43 µM $FeSO_4$ to the above medium (SI Tab. S2). Bottles of autoclaved Phytagel used for the above media mixtures were prepared by slowly adding weighed Phytagel powder to a beaker containing a suitable volume of room temperature water while rapidly stirring, to prevent formation of clumps. After further stirring, the beaker was covered in plastic wrap and microwaved until just boiling. After cooling the Phytagel, the solution was dispensed into bottles for autoclaving. The Phytagel media mixtures were dispensed using an Ovation ali-Q LS pipet controller (VistaLab) into sterilized borosilicate glass 16 × 150 mm tubes (DWK Life Sciences #73500-16150) covered with sterilized 16 mm OD polypropylene closures (DWK Life Sciences #73665-16). The bottom Glc-layer (typically with 2% (w/v) glucose and no cells) was first dispensed into the tubes and allowed to solidify at room temperature. Then, cells were quickly mixed into the (typically glucose-free) top layer Phytagel media solution in sterile 50 mL conical tubes, and this top C-layer (containing approx. 2E6 to 4E6 cells/mL) was dispensed on top of the solidified bottom layer. The Phytagel columns were inoculated with bacterial cells from overnight cultures grown in Luria-Bertani (LB) broth (10 g/L tryptone, 5 g/L yeast extract, 10 g/L NaCl) at 30°C, after 3000$g$ centrifugation to remove spent liquid media. Very similar patterns of bands form in columns inoculated with cells grown from LB, the synthetic media described above, or with exponentially growing (mid-log) or stationary phase cells. Therefore, inocula from LB overnight cultures was chosen for convenience, due to their faster growth and higher cell densities. In a typical column, the Glc-layer is 3.34 ± 0.10 mL and the C-layer is 10.02 ± 0.17 mL. Given the inner diameter of the tubes used (1.355 ± 0.015 cm), the height of the typical C-layer is 6.95 ± 0.19 cm. To minimize evaporation and gel shrinkage during the approx. 2 week growth period, while still allowing for sterile gas exchange at the top of the tube, the polypropylene closures were replaced by an assembly consisting of a size 0 silicone stopper (Cole-Parmer Instrument Co. #EW-06298-04) with a blunt 14G × 1.5 inch dispensing needle pushed through the stopper, and a 4 mm diameter, 0.22 um pore size, hydrophobic PTFE syringe filter (AQ Syringe Filters #358022-P04-C) fit in the needle's Luer port.

**Time-Lapse Imaging Setup** A custom tube rack for a row of twelve 16 × 150 mm tubes was laser cut and assembled from transparent acrylic sheets (McMaster-Carr #8560K239), to which a matte black background was added (either using foam board or spray-painted acrylic) for contrast during image acquisition. The tube rack was placed on top of an Artograph LightPad 920 LX LED Light Box, which was the sole light source for visualization. In later iterations of the setup, we were able to reduce reflections on the round glass tubes by placing 3M Privacy Filters (3M #PFTAP007) on top of the light box. Further improvements (not essential) could be made by inserting divider walls laser-cut from matte black acrylic sheets (Johnson Plastics Plus #311401V-QTR) between each tube in the holder. An open-ended box (open at the front to the camera) constructed from foam boards was placed to block stray ambient light from the back, left, top, and right. Two 80 mm USB-powered fans (Thermaltake) were placed diagonally at the left-front and right-front to circulate air inside the imaging area as a precautionary measure against temperature gradients. A DSLR camera (Canon) connected to an intervalometer was used to acquire images every 30 minutes, i.e. 48 images per day, with the entire setup at 30°C in an environmental chamber. Images were processed and analyzed as described in the SI Methods (SI §IA).

**RNA-seq Transcriptomics from Phytagel Columns** Genomic DNA-free and rRNA-depleted RNA, prepared from *S. marcescens* cells in bands recovered from liquefied Phytagel column slices (SI §I), were used in conjunction with a KAPA RNA HyperPrep kit (Roche KK8540), to create stranded RNA-seq libraries, which were Illumina-sequenced for 37 bp paired-end reads. Quality-control checked, filtered, and adapter-clipped sequencing reads were aligned to the ATCC 13880 genome, and the resulting (genes x samples) raw read count matrix was normalized and variance-stabilized via the rlog (regularized log) transform using DESeq2 [63] prior to PCA analysis. See SI Methods (SI §I) for further details.

## Appendix B: Mathematical formulation of model

The following partial differential equations (PDEs) implement the scheme depicted in Fig. 5a,

$$\mathrm{d}_t g(x,t) = D_g\, \partial_x^2 g(x,t) - y_g J_\mathrm{p}(x,t) \tag{B1a}$$

$$\mathrm{d}_t h(x,t) = D_h\, \partial_x^2 h(x,t) + y_h J_\mathrm{p}(x,t) - J_\mathrm{n}(x,t) \tag{B1b}$$

$$\mathrm{d}_t b(x,t) = D_b\, \partial_x^2 b(x,t) - J_\mathrm{n}(x,t) \tag{B1c}$$

$$\mathrm{d}_t s_0(x,t) = -J_\mathrm{r}(x,t) + J_\mathrm{q}(x,t) \tag{B1d}$$

$$\mathrm{d}_t s_1(x,t) = J_\mathrm{r}(x,t) - J_\mathrm{q}(x,t) + J_\mathrm{p}(x,t)\,. \tag{B1e}$$

Introducing a notation simpler than that used in Sec. V, the unknown functions represent, respectively, the concentration of glucose ($g$: Glc), protons ($h$: $H^+$), base molecules ($b$: B), es well as the density of cells in the quiescent and proliferating state ($s_i$: $S_i$ for $i = 1, 2$). These equations are abstractly written in terms of the rate at which cells transition to proliferation (break quiescence), $J_r(x,t)$, proliferate, $J_p(x,t)$, and revert to the quiescent state, $J_q(x,t)$. The term $J_n(x,t)$ represents the rate at which the base neutralizes protons, $H^+ + B \longrightarrow \varnothing$. The coefficients, $D_g$, $D_h$, and $D_b$ denote the diffusion coefficients of glucose, protons, and the base molecules, respectively, whereas the stoichiometric coefficients $y_g$ and $y_h$ characterize how much glucose is consumed and protons produced upon proliferation.

Notice from Eqs. (B1) that solely proliferation consumes glucose and produces protons. However, the concentrations of glucose and protons do affect the kinetics of transitions between cell states. Indeed, we use the following expressions of the rates of transition to proliferation, proliferation, and transition to quiescence, respectively,

$$J_r(x,t) = k_r\, s_0(x,t)\, \theta(g(x,t) - g^*)\, \theta(h^* - h(x,t)) \tag{B2a}$$

$$J_p(x,t) = k_p\, s_1(x,t)\, \theta(g(x,t) - g^*)\, \theta(h^{**} - h(x,t)) \tag{B2b}$$

$$J_q(x,t) = k_q\, s_1(x,t)\, [1 - \theta(g(x,t) - g^*)\, \theta(h^{**} - h(x,t))] \,, \tag{B2c}$$

where

$$\theta(X) = \begin{cases} 1 & \text{if } X \geq 0 \\ 0 & \text{otherwise} \end{cases} \tag{B3}$$

denotes the Heaviside step function. The coefficients $k_r$, $k_p$, and $k_q$, denote the rate constants of the respective processes. As introduced in Sec. V, the proton-concentration threshold above which cells stop proliferating, here denoted by $h^{**}$, is different from that aboee which cells stop transitioning to proliferation, $h^*$. We assume $h^* < h^{**}$, i.e. there exists a range of proton concentrations in which proliferating cells keep proliferating and quiescent cells remain quiescent. Reversion to quiescence occurs whenever cells stop proliferating. Transition to proliferation and proliferation itself also require glucose to be above a threshold, $g^*$, which we assume to be the same for the two processes. Finally, for the rate at which the base molecules neutralize protons, we assume mass action kinetics,

$$J_n(x,t) = k_n\, h(x,t)\, b(x,t) \,, \tag{B4}$$

where the coefficient $k_n$ denotes the corresponding rate constant.

We aimed at simulating the conditions of the kymographs in Fig. 3c. To do so, we used the following scheme of boundary and initial conditions.

The positions of the bottom *E*dge of the column, the *B*ottom interface (between the Glc- and C-layer), and the *T*op interface (between the C-layer and the air) are denoted by $x = x_E < 0$, $x = x_B = 0$, and $x = x_T > 0$, respectively. We denote by $\Delta x$ the height of the volume of solution containing the base added to the top interface. The boundaries are thus $x = x_E$ and $x = x_T + \Delta x$, and we impose the derivative of all unknown variables ($g, h, b, s_0$, and $s_1$) to vanish at these positions (Neumann boundary conditions). Regarding the initial conditions, glucose is initially placed in the Glc-layer,

$$g(x,0) = \begin{cases} g_0 & \text{if } x_E < x < 0 \\ 0 & \text{otherwise} \end{cases} \,, \tag{B5}$$

whereas cells in the C-layer [64], and we assume them to be in the quiescent state,

$$s_0(x,0) = \begin{cases} c_0 & \text{if } 0 < x < x_T \\ 0 & \text{otherwise} \end{cases} \tag{B6}$$

$$s_1(x,0) = 0 \quad \text{for all } x \,. \tag{B7}$$

The base molecules, instead, are initially located in the small volume added at the top interface,

$$b(x,0) = \begin{cases} b_0 & \text{if } x_T < x < x_T + \Delta x \\ 0 & \text{otherwise} \end{cases} \,, \tag{B8}$$

whereas the initial concentration of protons is constant throughout the system,

$$h(x,0) = h_0 \quad \text{for all } x \,. \tag{B9}$$

The ridge plots in Figs. 5b and 5c are obtained by numerically integrating the PDE model described in this section with parameters given in [43, Tab. S6]. Among all parameters, $k_r, k_q, k_n, y_h$, and $h^{**}$ have been treated as free parameters and the values reported in [43, Tab. S6] follow from a coarse parameter search. All the other parameters reflect values found for *E. coli*, known constants (such as diffusion coefficients), or experimental parameters (initial and boundary conditions). PDEs are implemented and integrated numerically using ModelingToolkit.jl [65] and DifferentialEquations.jl [66] (Julia language).

# Supplementary Information
# Bacterial proliferation pattern formation

John S. Chuang,[1,*] Riccardo Rao,[1,2,3,4,*] and Stanislas Leibler[1,2]

[1] *The Rockefeller University, Laboratory of Living Matter, New York, NY 10065, U.S.A.*
[2] *Institute for Advanced Study, Simons Center for Systems Biology, School of Natural Sciences, Princeton, NJ 08540, U.S.A.*
[3] *Medical University of Vienna, Center for Medical Data Science, Institute of the Science of Complex Systems, Spitalgasse 23, 1090 Vienna, Austria*
[4] *Complexity Science Hub, Josefstädter Strasse 39, 1080 Vienna, Austria*

## CONTENTS

| Strain | Information | Notes |
|---|---|---|
| ATCC 13880 | *S. marcescens* type strain | |
| DH5$\alpha$*pir* | DH5$\alpha$ *zdg-232::Tn10 uidA::pir*$^+$ | [1]; *E. coli pir*$^+$ host strain for R6K$\gamma$*ori* plasmids, derived from DH5$\alpha$ |
| MFD*pir* | MG1655 RP4-2-Tc::[$\Delta$Mu1::*aac(3)IV-$\Delta$aphA-$\Delta$nic35-$\Delta$Mu2::zeo] $\Delta$dapA::(erm-pir) $\Delta$recA* | [2]; Mu Free Donor *E. coli* strain, derived from MG1655 |
| JCSm010 | ATCC 13880 $\Delta$*asnB* | This work; knockout of GSMA_04566 |
| JCSm022 | ATCC 13880 $\Delta$*sucD* | This work; knockout of GSMA_04522 |
| JCSm026 | ATCC 13880 $\Delta$*slaAB* | This work; knockout of GSMA_02274 and GSMA_02275 |
| D1 | *S. marcescens* stably red-pigmented strain | |
| WCF | *S. marcescens* colorless mutant | |

TABLE S1: **Strains.**

---

* These authors contributed equally to this work.

| medium name | composition |
|---|---|
| synthetic medium 1 | 4.4 mM $K_2HPO_4$, 0.2 mM $MgSO_4$, 25 µM $MnCl_2$, 7 µM $CaCl_2$, 7 µM $FeSO_4$ |
| synthetic medium 2 | synthetic medium 1 + 5 µM $Na_2MoO_4$, 5 µM $H_2SeO_3$, and an additional 43 µM $FeSO_4$ |

TABLE S2: **Mineral nutrient composition of growth media (prior to carbon and nitrogen source addition)**. Note that the additional 4 mM $MgSO_4$ used for Phytagel gelation are not accounted for in the values above.

# I. EXTENDED METHODS

## A. Image Analysis of Phytagel Columns

All camera images were converted to grayscale for image analysis. For each experiment, the calculated median angle of all near-vertical lines (±2.5° from vertical) detected by Canny edge and Hough transform of the region containing the test tubes in the first image ($t = 0$) was used to rotate all subsequent images. Rectangular regions-of-interest (ROIs) along the length of each tube were created, making sure to avoid the walls of the tube. For each experiment, the first image (prior to bacterial growth) was designated as the background image, and each subsequent image was background-subtracted and the resulting image intensities along the tube length in each ROI were quantified, averaging across each ROI's width. The resulting intensity data for each tube at different times was used to construct the kymograph heat maps.

To identify the positions and times of appearance of bands, the following procedure was performed. First, the signals from the intensity data were smoothed in space (pixels) and time (frames) via a running median. Second, band positions were identified by assessing the peak of light intensity using the positions of the bottom and top interfaces (which are similarly identified as peaks of light intensity) as references. Since the height of the typical C-layer is 6.95 ± 0.19 cm, points at distance $x$ (in cm) from the bottom interface have an uncertainty equal to $x \cdot 0.19/6.95$ cm. Third, band appearance times were recorded as the times at which the intensity of light at a band position reaches a certain threshold (the precise value depends on the specific experiment, since different experiments may be performed with different camera set-ups). The uncertainty on band appearance time corresponds to the temporal resolution of our set up, *viz.* 30 minutes.

## B. RNA-seq Transcriptomics from Phytagel Columns

RNA-seq transcriptomics analysis of different bands was performed using Phytagel columns under modified media conditions and geometry. Instead of the closed-ended test tubes of the standard time-lapse growth experiments, open-ended Pyrex tubing (Wale Apparatus Co. #BS-025) cut to 170 mm length with fire-polished ends were used to allow easy removal of the gel column for destructive band excision. Furthermore, the tubing diameter was larger (25 mm OD; 22 mm ID) in order to obtain sufficient cells from a band for RNA isolation. Size #3 neoprene stoppers (Cole-Parmer #EW-62991-10) were used to seal the bottom of the tubing, and the top was covered by a polypropylene closure (Kimble #73662-25). This assembly was sterilized by autoclaving. Phytagel columns for the RNA-seq experiment consisted of a 20 mL Glc-layer containing 1.4% (w/v) glucose and a 20 mL top C-layer containing cells. The media for the RNA-seq experiment used a mixture of 15 L-amino acids (instead of just 6.8 mM glutamate) composed of 0.66 mM Arg, 0.58 mM His, 1.33 mM Ile, 2.61 mM Leu, 2.02 mM Lys, 0.67 mM Met, 0.42 mM Phe, 1.57 mM Ser, 1.49 mM Thr, 1.94 mM Val, 3.4 mM Glu, 1.7 mM Asp, 1.3 mM Gly, 0.06 mM Tyr, 0.08 mM Cys), and Phytagel was reduced to 0.2% (w/v).

At several time points (i.e. after the formation of a new band), three columns were destructively sampled by removing the bottom stopper and sliding the gel onto plastic wrap, where gel slices containing bands were excised with a clean razor blade, transferred into a 50 mL conical tube, and quickly broken up using a plastic rod (e.g. Nunc #251586) to increase surface area. Immediately after, 12 mL 5 mM EDTA in STOP solution (10% (v/v) EtOH, 1% (v/v) water-saturated phenol; [3]) was added, and the tube was vigorously vortexed for 2 minutes until Phytagel liquefication to release the cells from the gel. The cells were centrifuged at 5 000 $g$ for 20 minutes, resuspended in 1 mL EDTA/STOP, transferred to a 1.7 mL microcentrifuge tube, centrifuged at 5 000 $g$ for 5 minutes to remove all liquid, and the cell pellet was snap-frozen and stored at −80°C.

All RNA isolation and manipulation steps were performed in an separate workspace with RNase-free chemicals and plasticware, with all working surfaces wiped clean with RNaseZAP (Sigma). Total RNA was prepared using the RNAsnap method for Gram negative bacteria [4], followed by RNA Clean & Concentrator-5 (CC-5) column purification (Zymo Research) using the >200 nt protocol variant, including the manufacturer's optional in-column DNaseI digestion step. After elution, the RNA was further digested with 1 µL TURBO DNase (Invitrogen) at 37°C for 30 minutes. Then an additional 1 µL TURBO DNase was added and the reaction was allowed to continue for another 30 minutes. The 50 µL reactions were diluted with water to 150 µL and purified on a *new* CC-5 column and eluted in 20 to 40 µL. Since ribosomal RNA depletion in the next step does not select against genomic DNA (gDNA), these important steps ensured against gDNA contamination, which was verified by 30 cycles of PCR with a pair of ATCC 13880 primers (13880_fwd1 and 13880_rev1) in parallel with a control sample which had not been digested with TURBO DNase. The RNA was checked by Qubit RNA HS (Thermo Fisher Scientific), Bioanalyzer (Agilent), and by RNA bleach gels [5] containing 1.4% agarose, 1% (v/v) commercial Chlorox in 0.5X TBE with GelRed (Biotium) added to the gel.

Ribosomal RNA was depleted using Ribo-Zero (Illumina), checked by Qubit RNA HS and Bioanalyzer, and then used to prepare stranded RNA-seq libraries using a KAPA RNA HyperPrep kit (Roche KK8540) with KAPA Dual Index adapters (Roche KK8722). The resulting libraries were checked on a 4% Agarose-1000 (Thermo Fisher Scientific) gel, quantified with Qubit dsDNA HS, and then pooled for Illumina sequencing (NextSeq 500 High Output) for 37 bp paired-end reads.

### C. Bioinformatics

FASTQ files were quality-checked with FastQC [6], filtered with AfterQC [7], preprocessed with Trimmomatic [8] to remove any low quality leading or trailing bases and for adapter clipping, and then aligned using Bowtie [9] with the ATCC 13880 genome ([10]; NCBI Accession #JOVM00000000, downloaded from [11]). The resulting BAM files were sorted using SAMtools [12]. Read counts for different annotated gene features were determined using featureCounts [13] against a GTF file created by first downloading an annotated GFF3 file [14] from Ensembl Bacteria, Release 42 [15] and then converting it to GTF using rtracklayer [16]. The raw read count matrix was first normalized and then variance-stabilized (with the regularized log transform) using DESeq2 [17] prior to PCA analysis.

### D. Mutant Strain Construction

We used the pTOX allelic exchange vector system [18] to construct scarless deletion mutants of *S. marcescens*. The pTOX6 vector (AddGene #127451) contains a positive selection marker ($Cm^R$), a negative selection marker (toxin Tse2 induced by rhamnose and repressed by glucose), the AmilCP blue chromoprotein from the coral *Acropora millepora*, a defective R6K origin of replication which can only replicate in *pir*-containing hosts (such as DH5$\alpha$*pir* and MFD*pir*), and a conjugative mobilization region mobRP4 which can be mobilized from donor hosts carrying transfer genes from the broad host range plasmid RP4.

The following abbreviations and media formulations are used below: Cm = chloramphenicol; Cm20 = 20 µg/mL Cm; Cm40 = 40 µg/mL Cm; Cm300 = 300 µg/mL Cm; Glc = D-glucose (aka Dextrose); Glc2 = 2% (w/v) Glc; Rha = Rhamnose; Rha2 = 2% (w/v) Rha; DAP = 0.3 mM diaminopimelic acid; LB = 10 g/L tryptone, 5 g/L yeast extract, 10 g/L NaCl; M9 = M9 Minimal Salts (Difco) + 0.5 mM $MgSO_4$ + 0.1 mM $CaCl_2$ + 25 µM $FeCl_3$ (from 50 mM stock in 100 mM citrate) + 0.2% (w/v) Casamino acids. For agar plates, Difco agar was used at 1.5% (w/v), autoclaved separately from phosphates.

For each gene $X$ to be knocked out, three gene $X$ specific primer pairs were designed. Primer Pair 1 ($X$_up_FWD and $X$_up_REV) and Primer Pair 2 ($X$_down_FWD and $X$_down_REV) were designed using the NEBuilder HiFi Assembly module of SnapGene software (GSL Biotech, LLC) to join the $\sim$ 700 bp regions immediately upstream ($X$_up) and downstream ($X$_down) of the gene $X$ coding sequence (CDS), into pTOX6 cut with XhoI and NheI, resulting in plasmid pTOX6-KO($X$). Primer Pair 3 ($X$_check_up and $X$_check_down) is located in the genome approx. 100 to 200 bp outside of the homology regions. Primers were ordered from Integrated DNA Technologies.

pTOX6 plasmid, a gift from Matthew Waldor (Addgene plasmid # 127451; [19]; RRID:Addgene_127451), was purified (QIAprep Spin Miniprep Kit; Qiagen) after growing its DH5$\alpha$*pir* host in LB/Cm20/Glc2 broth. pTOX6 was digested with restriction enzymes (New England Biolabs) XhoI and NheI-HF at 37°C, heat inactivated at 80°C for 10 minutes, and purified (QIAquick PCR Purification Kit; Qiagen). ATCC 13880 genomic DNA (gDNA) was prepared using the GenElute Bacterial Genomic DNA Kit (Sigma).

For each target gene $X$, the upstream ($X$_up) and downstream ($X$_down) homology regions were amplified from ATCC 13880 gDNA using Q5 Hot Start High Fidelity DNA Polymerase (New England Biolabs), checked on an agarose gel, cleaned up (QIAquick PCR Purification Kit; Qiagen), and quantified using Qubit dsDNA HS (Thermo Fisher Scientific). pTOX6-KO(X) was then assembled at 50°C for 1 hour in 10 µL NEBuilder HiFi DNA Assembly (New England Biolabs) reactions containing 0.01 pmol digested pTOX6 vector, 0.02 pmol $X$_up PCR product, and 0.02 pmol $X$_down PCR product, checked on agarose gels, and transformed into MFD*pir* (CRBIP 19.334), which was obtained from the Collection de l'Insitut Pasteur (CRBIP, Paris, France).

MFD*pir* competent cells were prepared using Zymo Mix & Go (Zymo Research) after growth in ZymoBroth + DAP. The optional heat shock step (42°C for 45 seconds, then quenched on ice) was critical for obtaining decent transformation efficiency for this strain. Cells were recovered in S.O.C. medium (Thermo Fisher Scientific) supplemented with DAP plus an extra 1.5% Glc at 37°C for 1 hour, and then plated on LB/DAP/Cm40/Glc2. Resulting colonies were restreaked for colony purification, and PCR-tested with primer pairs which tested for i) the presence of an insertion in pTOX6 (using primers pTOX6_check_up and pTOX6_check_down) and ii) the formation of the correct $X$_up and $X$_down junctions with pTOX6.

For conjugation, the $Dap^-$, $Cm^R$ *E. coli* donor strains (MFD*pir* harboring pTOX6-KO(X) plasmids) were grown in LB/-DAP/Cm40/Glc2 broth. The $Dap^+$, $Cm^S$ *S. marcescens* acceptor strain ATCC 13880 was grown in LB. Cells were centrifuged and washed twice in phosphate-buffered saline (PBS), and for each conjugation, a 50:1 (v/v) donor:acceptor mixture was spotted on LB/DAP/Glc2 plates (no Cm) at 30°C overnight, before scraping and resuspending the cells in PBS, and plating on LB/Glc2/Cm300 (no DAP) to select for recipient *S. marcescens* cells with genome-integrated pTOX6-KO($X$). After restreaking, colony PCR using two primer sets - i) $X$_check_up + $X$_down_REV and ii) $X$_check_down + $X$_up_FWD – was performed to test candidate transconjugant clones for the formation of new junctions in the genome between the integrated pTOX6_KO($X$) construct and both sides of gene $X$. To select for loopout events (which can either revert to wild-type or knock out gene $X$), transconjugants were diluted 1:50 into LB/Glc2 and grown for 3 hours at 30°C, washed twice with M9 Salts/Rha2 (to induce the counterselectable Tse2 toxin marker from pTOX6), and then plated and restreaked on M9/Rha2 plates, and finally tested by colony PCR using $X$_check_up and $X$_check_down to distinguish between true knockout versus reversion events. The resulting scarless knockout mutant strains could subsequently be stably propagated on LB.

### E. Primer Sequences

Primer sequences are reported in Tab. S3. The ATCC 13880 genome [10] used for primer design was based on annotations from genome assembly GSMA_DRAFTv1, accessed at [20] from Ensembl Bacteria, Release 45 [15]. Annotated genes have GMSA_*nnnnn* IDs, where asnB = GSMA_04566, fliC = GSMA_02837, nuoA = GSMA_02408, pflB = GSMA_04054, slaA = GSMA_02274, slaB = GSMA_02275, and sucD = GSMA_04522.

## II. REPRODUCIBILITY

In this section, we discuss the reproducibility of PPF in greater depth.

Figure S1a shows the kymographs of a 12-replicate experiment in which the medium contains solely glutamate as nitrogen source (Mov. S1). Figure 1 in the main text shows the dynamics of the 7th column, whereas Fig. 2 (main text) that of columns 7–9.

To assess the reproducibility of the PPF dynamics, we plot the standard deviation (SD) of band coordinates (position and time of appearance) versus their average values using different replicates as samples (Figs. S1c and S1d). As shown Fig. S1c, the SDs of band positions are all smaller than the measurement uncertainty, which makes the pattern shapes statistically indistinguishable across replicates. In contrast, band appearance times are distinguishable and their SDs increase roughly exponentially going from the first to the last band.

To test the pattern formation reproducibility under small variations in the experiment set up, we conducted an independent 12-replicate experiment (Fig. S2a), where each pair of columns contains cells from one of three different *S. marcescens* colonies growing in one of two separate preparations of media type 1 (Tab. S2). Also in this case, pattern shapes are indistinguishable (Fig. S2c), and band appearance times grow approx. exponentially (Fig. S2d).

## III. SCALING

In this section, we discuss the scaling law of band coordinates for the experiment in Fig. S1a. To do so, we rewrite the relation introduced in the main text

$$\log x_{ij} = \frac{1}{2}\log K + \frac{\alpha}{2}\log\left(t_{ij} - \tau\right) + \epsilon_{ij}\,, \tag{S1}$$

where $(x_{ij}, t_{ij})$ are the spatiotemporal coordinates of the $i$-th band for the $j$-th replicate, $K$ and $\alpha$ the anomalous diffusion coefficient and exponent, $\tau$ a time shift, and $\epsilon_{ij}$ an error term. For this last term, we assume it to be white noise, i.e. independently, identically, and normally distributed with zero mean and constant variance. The parameters, $\alpha$, $K$, and $\tau$ are estimated via maximum likelihood (ML) using the implementation developed in the Julia packages Turing.jl [21] and Optim.jl [22]. The Bayesian Information Criterion (BIC) and the Coefficient of Determination ($r^2$) support the inclusion of the time shift $\tau$ in the model (Tab. S4). In Fig. S1b, we compare the two models on a log–log scale. When either model is chosen, the scaling relation is incompatible with canonical diffusion, and suggests a sub-diffusive scaling: from the ML estimates of $\alpha$ and their confidence interval, the probability that $\alpha \approx 1$ is $p \ll 1$ (Tab. S4).

To clarify how the pattern and its scaling law respond to changes of initial glucose concentration, consider the experiment whose kymographs are plotted in Fig. S3. In this experiment, the C-layer is larger (total height equal to 9.3 ± 0.2 cm), the Glc-layer smaller (volume approx. 1 $cm^3$), and each group of 4 columns is prepared with a different glucose concentration (6, 8, and 10 % (w/v)). Figure 2c in the main text reproduces the 2nd, 7th, and 11th column of Fig. S3, whereas the squared of band positions are plotted vs their appearance time in Fig. 2d (main text).

Based on evidence from this experiment, we argue that the scaling law approaches that of canonical diffusion as the concentration of glucose increases. Table S5 compares the outcome of fitting the data using three possible relationships. The first one is based on (S1) and the assumption that the scaling relation is not affected by the concentration of glucose. The second relationship assumes that solely the anomalous diffusion coefficient changes with varying glucose, and it does so in a linear fashion:

$$\begin{aligned} \log x_{ij} &= \frac{1}{2}\log K_j + \frac{\alpha}{2}\log\left(t_{ij} - \tau\right) + \epsilon_{ij} \\ K_j &= K_0 + K_1[\mathrm{Glc}]_j/[\mathrm{Glc}]_1\,, \end{aligned} \tag{S2}$$

where $K_0$, $K_1$, $\alpha$, $\tau$, and $\sigma$ (the noise standard deviation) are the fitting parameters, $[\mathrm{Glc}]_j$ is the initial concentration of glucose in the column $j$, and $[\mathrm{Glc}]_1$ an arbitrary reference value set to 1 % w/v. Finally, the third relationship assumes that scaling exponent changes linearly:

$$\begin{aligned} \log x_{ij} &= \frac{1}{2}\log K_j + \frac{\alpha_j}{2}\log\left(t_{ij}-\tau\right) + \epsilon_{ij} \\ \alpha_j &= \alpha_0 + a_1[\mathrm{Glc}]_j/[\mathrm{Glc}]_1 \\ K_j &= K_0\left(\alpha_0 + \alpha_1[\mathrm{Glc}]_j/[\mathrm{Glc}]_1\right) \equiv K_0\,\alpha_j \ . \end{aligned} \tag{S3}$$

where the fitting parameters are $K_0$, $\alpha_0$, $\alpha_1$, $\tau$, and $\sigma$.

| primer name | sequence |
|---|---|
| 02408_up_FWD | TAAATGCATCCCGGGACGTCggatttaagatgctgggtttgtttttg |
| 02408_up_REV | atgcctcgctgtaagccgcaaagcccgatagc |
| 02408_down_FWD | ttgcggcttacagcgaggcattaagatggac |
| 02408_down_REV | AACAGGACACTTGGTATACGTGgcactaagatggacgttaaatgaacgt |
| 02408_check_down | ctggtcaggtttaacccataccacg |
| 02408_check_up | cagcaacgaactggccatgg |
| | |
| 02837_up_FWD | TAAATGCATCCCGGGACGTCccttcagcgcctaatttagcg |
| 02837_up_REV | ggagagacgatgtttgctttccttacgagtcag |
| 02837_down_FWD | aagcaaacatcgtctctcccgattaactgtc |
| 02837_down_REV | AACAGGACACTTGGTATACGTGgctgttattggtgtccaacaga |
| 02837_check_down | cggcgccgatctctttgag |
| 02837_check_up | tgttttcgccggtggtggaa |
| | |
| 04054_up_FWD | TAAATGCATCCCGGGACGTCcatctgcttctctctcgggttaatg |
| 04054_up_REV | gtcggcgttatgtacacctacctttgattgtggatttct |
| 04054_down_FWD | taggtgtacataacgccgaccgggc |
| 04054_down_REV | AACAGGACACTTGGTATACGTGtggtggatctcgtcgttcatctgt |
| 04054_check_down | cgctatccatttgtgcttgcc |
| 04054_check_up | gcgttgttttaacccttaagagcg |
| | |
| 04522_up_FWD | TAAATGCATCCCGGGACGTCagggccgtgagctgg |
| 04522_up_REV | actaggttcttagatggacattatttaccctccactgc |
| 04522_down_FWD | ataatgtccatctaagaacctagtacgaaataacaatatctcgacatg |
| 04522_down_REV | AACAGGACACTTGGTATACGTGcacctacagctttaacccggcg |
| 04522_check_down | ggacgtcggcgacatcaagg |
| 04522_check_up | attcatggcgtctaccgaaggc |
| | |
| 04566_up_FWD | TAAATGCATCCCGGGACGTCcgctgacggtctccgagct |
| 04566_up_REV | cgccgcgcgaacaactctcctaacgggcttttatg |
| 04566_down_FWD | ggagagttgttcgcgcggcgttttc |
| 04566_down_REV | AACAGGACACTTGGTATACGTGtggtggctacgacgggatttg |
| 04566_check_down | ggctgaacgccagacagcaa |
| 04566_check_up | caagagcaaagccggcgtg |
| | |
| slaAB_up_FWD | TAAATGCATCCCGGGACGTCcgtgaataccggtaatgtccgcca |
| slaAB_up_REV | cgatgatgaaagctgactcctccaccagc |
| slaAB_down_FWD | aggagtcagctttcatcatcgcgctgtttcacc |
| slaAB_down_REV | AACAGGACACTTGGTATACGTGatcgcgtcggcatattccg |
| slaAB_check_down | tctggtactgcaccagctgg |
| slaAB_check_up | ggcagcacataggttttgccg |
| | |
| pTOX6_check_up | ttttggcggatgagagggta |
| pTOX6_check_down | ccgatcaacgtctcattttcgcc |
| | |
| 13880_fwd1 | ccgtgaattagcaaagccgt |
| 13880_rev1 | agacgctaaacccatggtgg |

TABLE S3: **Primer sequences.** Capitalized bases in $X$_up_FWD and $X$_down_REV primers correspond to homology regions in pTOX6.

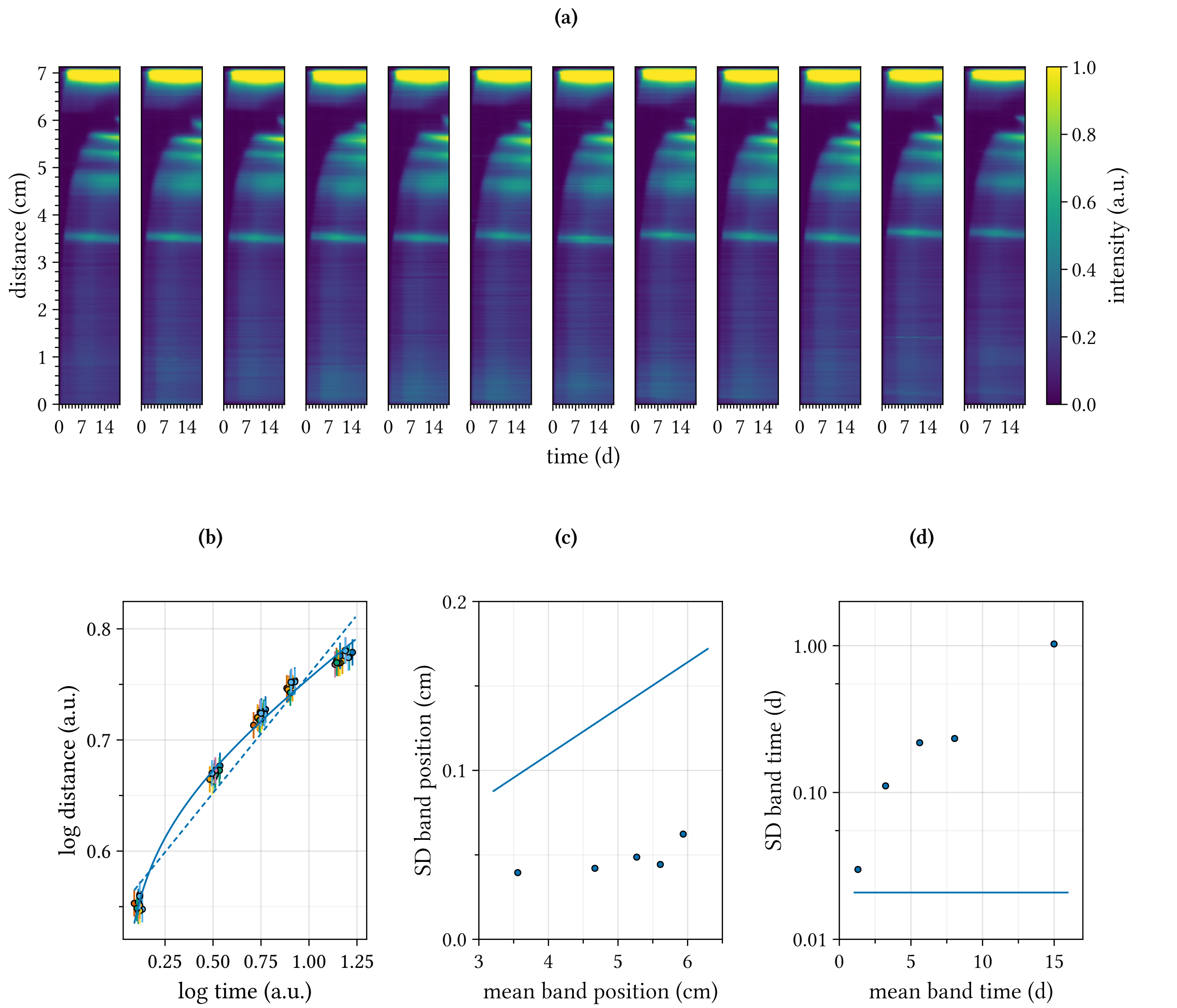

FIG. S1: **12-replicate experiment**. Medium composition: synthetic medium 1 (Tab. S2) + 6.8 mM glutamate. **(a)** Kymographs. It can be noticed that bands appear to shift downward as time increases. In part, this could be due to bacterial proliferation being activated at later times by the gradient of ammonia diffusing from the top interface. Although limited (see Methods), evaporation can also play a minor role. However, our estimates show that the downward shift is typically smaller than the corresponding errors (ranging from 1.3 to 2 mm depending on the position of the band), which clarify that this effect is minor compared to the main phenomenon of band formation. **(b)** Scaling of band appearance coordinates on log–log scale (base-10 logarithms). Dots and error bars represent band coordinates and their measurement uncertainties. The solid and dashed lines represent the relation in Eq. (S1) with or without the time shift, respectively. The parameters are estimated via maximum likelihood (Tab. S4). **(c)** Standard deviations of band position vs means of band position. The solid blue line represents the measurement uncertainty (§I A). **(d)** Standard deviations of band appearance time vs means of band appearance time. The solid blue line represents the measurement uncertainty (§I A).

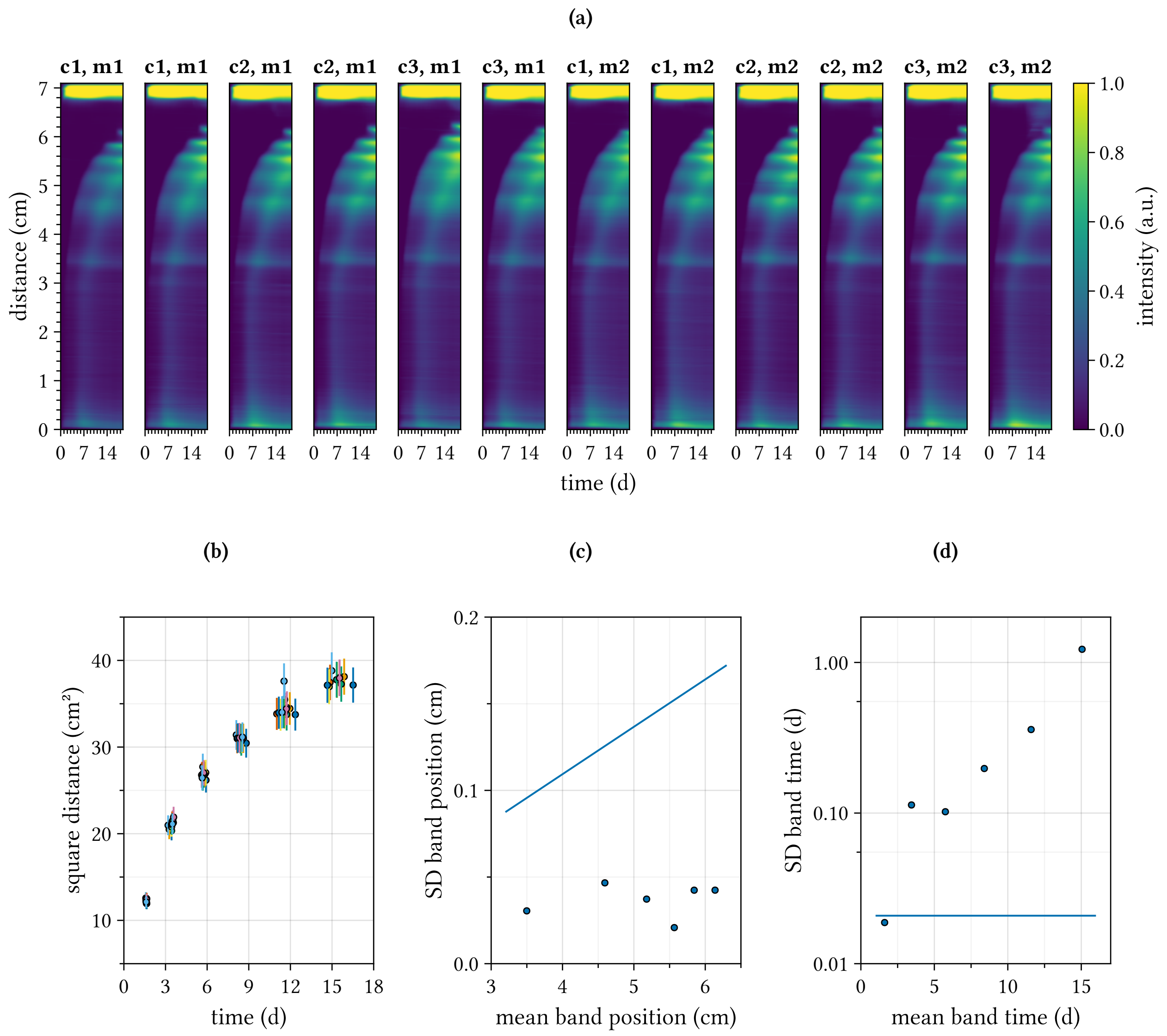

FIG. S2: **12-replicate experiment: three different colonies in two separate preparations of media.** Medium composition: synthetic medium 1 (Tab. S2) + 6.8 mM glutamate. **(a)** Kymographs. "c1, c2, c3" label columns inoculated with cells from one of three different *S. marcescens* colonies, whereas "m1, m2" label columns containing one of two separate preparations of the same media recipe. **(b)** Scaling of band appearance coordinates. Error bars represent measurement uncertainty. **(c)** Standard deviations of band position vs means of band position. The solid blue line represents the measurement uncertainty (§I A). **(d)** standard deviations of band appearance time vs means of band appearance time. The solid blue line represents the measurement uncertainty (§I A).

The third model outperforms the first two according to all model selection criteria used, and it shows that the scaling law

| **model** | **parameters** | **BIC** | $r^2$ |
|---|---|---|---|
| sub-diffusive w/out time shift | $\alpha = 0.43 \pm 0.01$,<br>$K = 12.3 \pm 0.1\,\mathrm{cm}^2\mathrm{d}^{-\alpha}$ | −300 | 0.95 |
| sub-diffusive w/ time shift | $\alpha = 0.27 \pm 0.01$,<br>$K = 18.1 \pm 0.3\,\mathrm{cm}^2\mathrm{d}^{-\alpha}$,<br>$\tau = 1.03 \pm 0.03\,\mathrm{d}$, | −410 | 0.99 |

TABLE S4: **Models of scaling law**. BIC, Bayesian Information Criterion; $r^2$, Coefficient of determination. From estimation of the confidence intervals of $\alpha$ for the relation without time shift (the first model), one excludes canonical diffusion ($\alpha \approx 1$): the probability that $\alpha > 0.6$ is $p < 10^{-16}$.

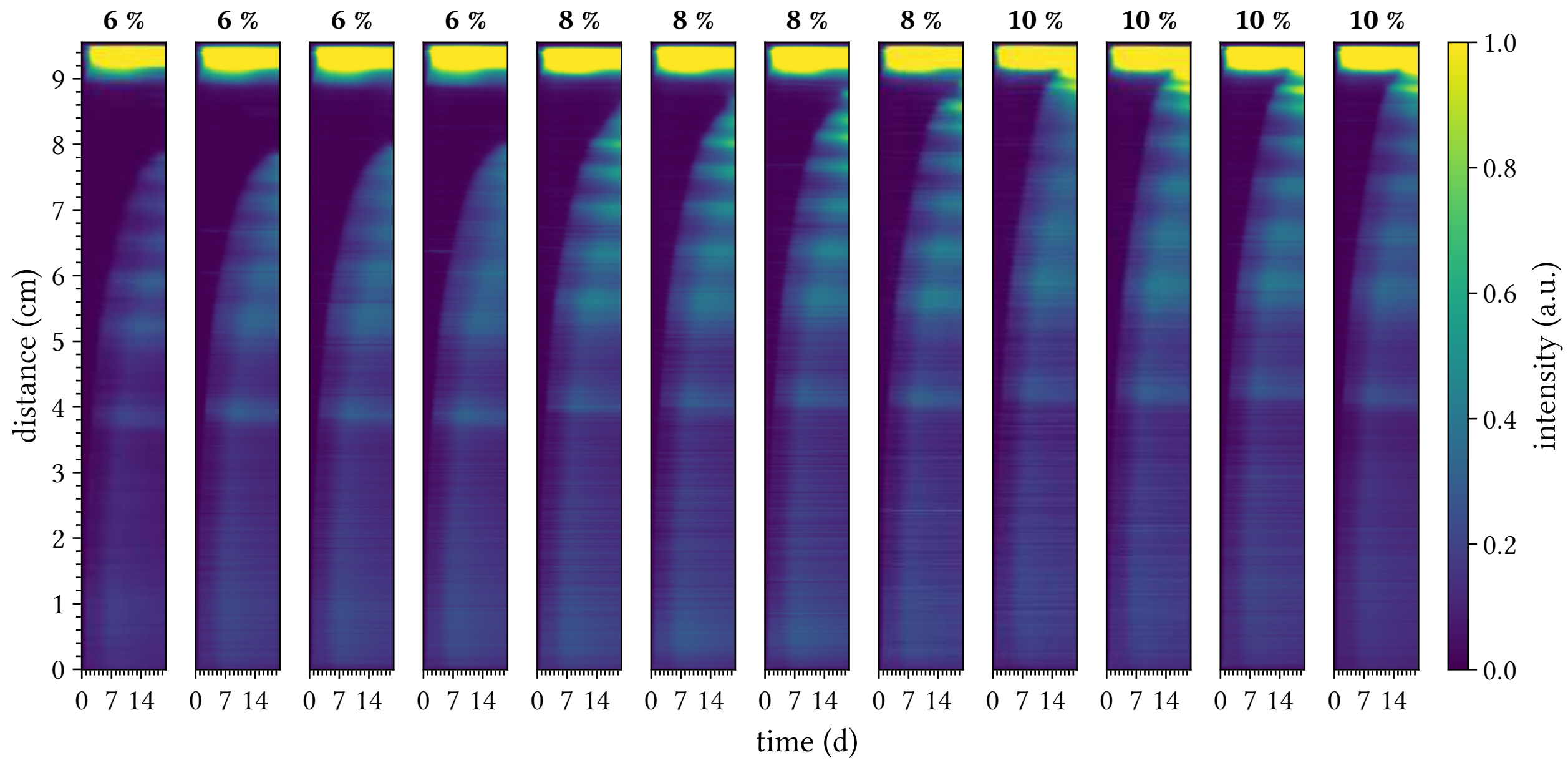

FIG. S3: **Titration of glucose**. Medium composition: synthetic medium 1 (Tab. S2) + 6.8 mM glutamate. The labels above each column denote the initial concentration of glucose in Glc-layer. Importantly, the geometry of these columns is different from the others: the volume of the Glc-layer is smaller (1 mL), that of the C-layer is larger (13.4 ± 0.2 mL), and hence the height of the latter is 9.3 ± 0.2 cm.

approaches the canonical diffusion scaling as the initial glucose concentration increases:

$$\frac{\mathrm{d}\alpha}{\mathrm{d}[\mathrm{Glc}]} = \frac{\alpha_1}{[\mathrm{Glc}]_1} = 0.020 \pm 0.001\ [\%\ (\mathrm{w/v})]^{-1} > 0 \tag{S4}$$

(95%-confidence intervals: $[0.017, 0.022]$).

## IV. ON THE ROLE OF AMINO ACID CATABOLISM

In §III (main text), we noticed that the amino acid catabolism operated by cells at the top interface entails the release of ammonia. Ammonia diffuses down the tube, where it seems essential for the formation of the top bands. This may be the case for two reasons: *(i)* ammonia enables cells in lower parts of the column to grow faster than amino acids would allow for, as the former is more easily taken up by the cells [23]; *(ii)* being a weak base, $NH_3$ neutralizes the acids released by glucose fermentation. We argue that the latter effect is dominant: *(i)* titrating ammonium ions (specifically $NH_4Cl$) to a medium prepared with a single amino acid has minor effects on the formation of the top bands (Fig. S4); *(ii)* as mentioned in the main text (Fig. 3c, main text), completely replacing the amino acids in the medium with ammonium ions disrupts the formation of top bands (Fig. S5a, 1st and 2nd columns). And yet their formation can be restored upon exogenous addition of phosphate buffer at the top interface (Fig. S5a, 3rd to 12th columns).

| model | parameters | BIC | $r^2$ |
|---|---|---|---|
| constant scaling, Eq. (S1) | $\alpha = 0.47 \pm 0.03$,<br>$K = 21.3 \pm 0.7\,\mathrm{cm^2 d^{-\alpha}}$,<br>$\tau = 1.5 \pm 0.1\,\mathrm{d}$ | −346 | 0.94 |
| changing an. diff. coef. Eq. (S2) | $\alpha = 0.51 \pm 0.02$,<br>$K_0 = 9.0 \pm 0.6\,\mathrm{cm^2 d^{-\alpha}}$, $K_1 = 1.37 \pm 0.07\,\mathrm{cm^2 d^{-\alpha}}$,<br>$\tau = 1.43 \pm 0.06\,\mathrm{d}$ | −494 | 0.99 |
| changing scaling exp. Eq. (S3) | $\alpha_0 = 0.37 \pm 0.02$, $\alpha_1 = 0.020 \pm 0.001$,<br>$K_0 = 36 \pm 2\,\mathrm{cm^2 d^{-\alpha}}$,<br>$\tau = 1.35 \pm 0.06\,\mathrm{d}$ | −505 | 0.99 |

TABLE S5: **Model of scaling law with response to changes of initial glucose concentration**. BIC, Bayesian Information Criterion; $r^2$, coefficient of determination.

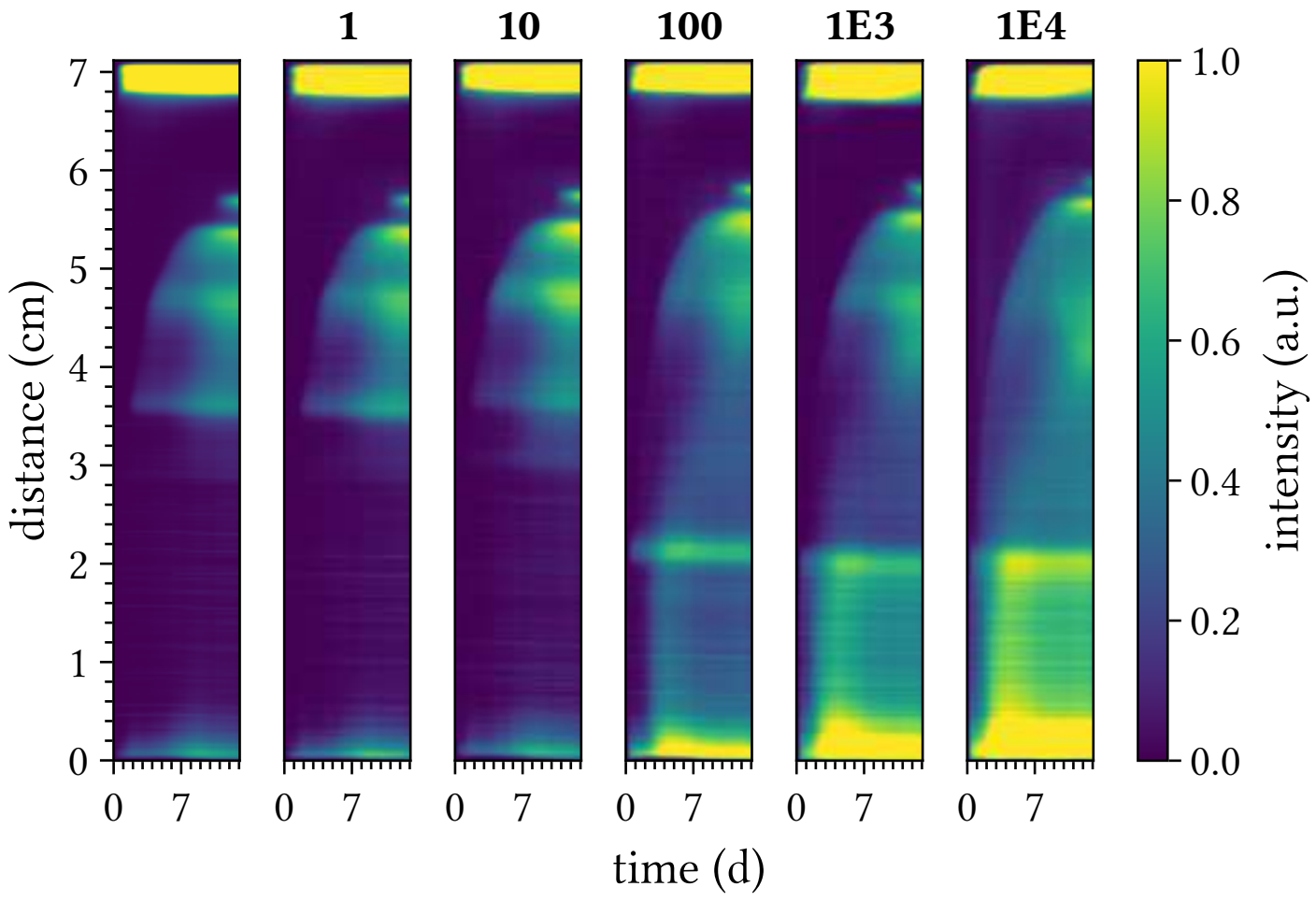

FIG. S4: **Titration of ammonium chloride in glutamate columns**. Medium composition: synthetic medium 2 (Tab. S2) + 6.8 mM glutamate + varying amount of $NH_4Cl$. The amount of $NH_4Cl$ in each column corresponds to 1.87 μM times the value reported on top of the column: none, 1.87 μM, 18.7 μM, 187 μM, 1.87 mM, and 18.7 mM. Notice how ammonium chloride affects the position of the bottom band, which sharply transitions from ∼ 3.5 cm to ∼ 2.1 cm. In contrast, top bands are relatively unaffected.

The experiment presented in Fig. S5 also clarifies that the base diffusing from the top interface significantly contributes to make the scaling sub-diffusive. To do so, we fit the band appearance coordinates using Eq. (S3) with [Glc] replaced by the buffer concentration, [K-phos7]. ML estimation ($r^2 = 0.99$) confirms that $\alpha$ significantly decreases with [K-phos7],

$$\frac{\mathrm{d}\alpha}{\mathrm{d[K\text{-}phos7]}} = \frac{\alpha_1}{\mathrm{[K\text{-}phos7]_1}} = -0.08 \pm 0.02\,\mathrm{M^{-1}} < 0 \tag{S5}$$

(95%-confidence intervals: $[-0.11, -0.05]$).

**(a)**

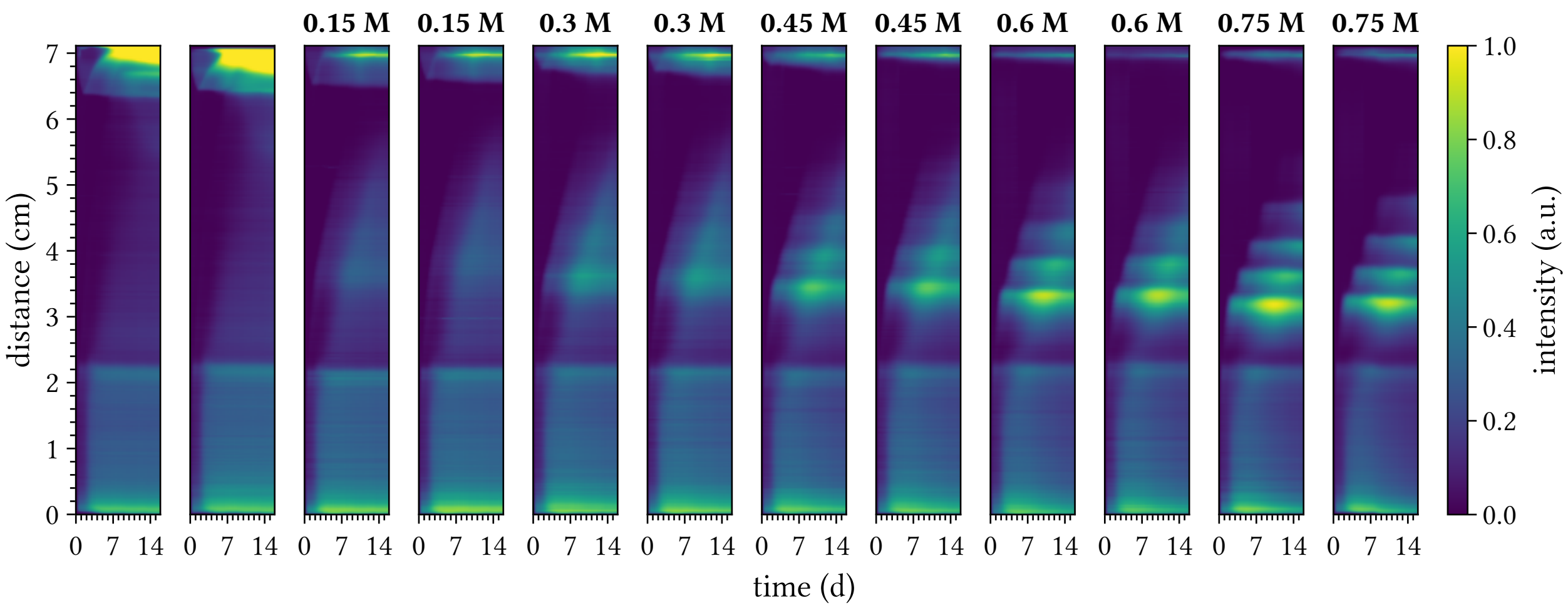

**(b)**

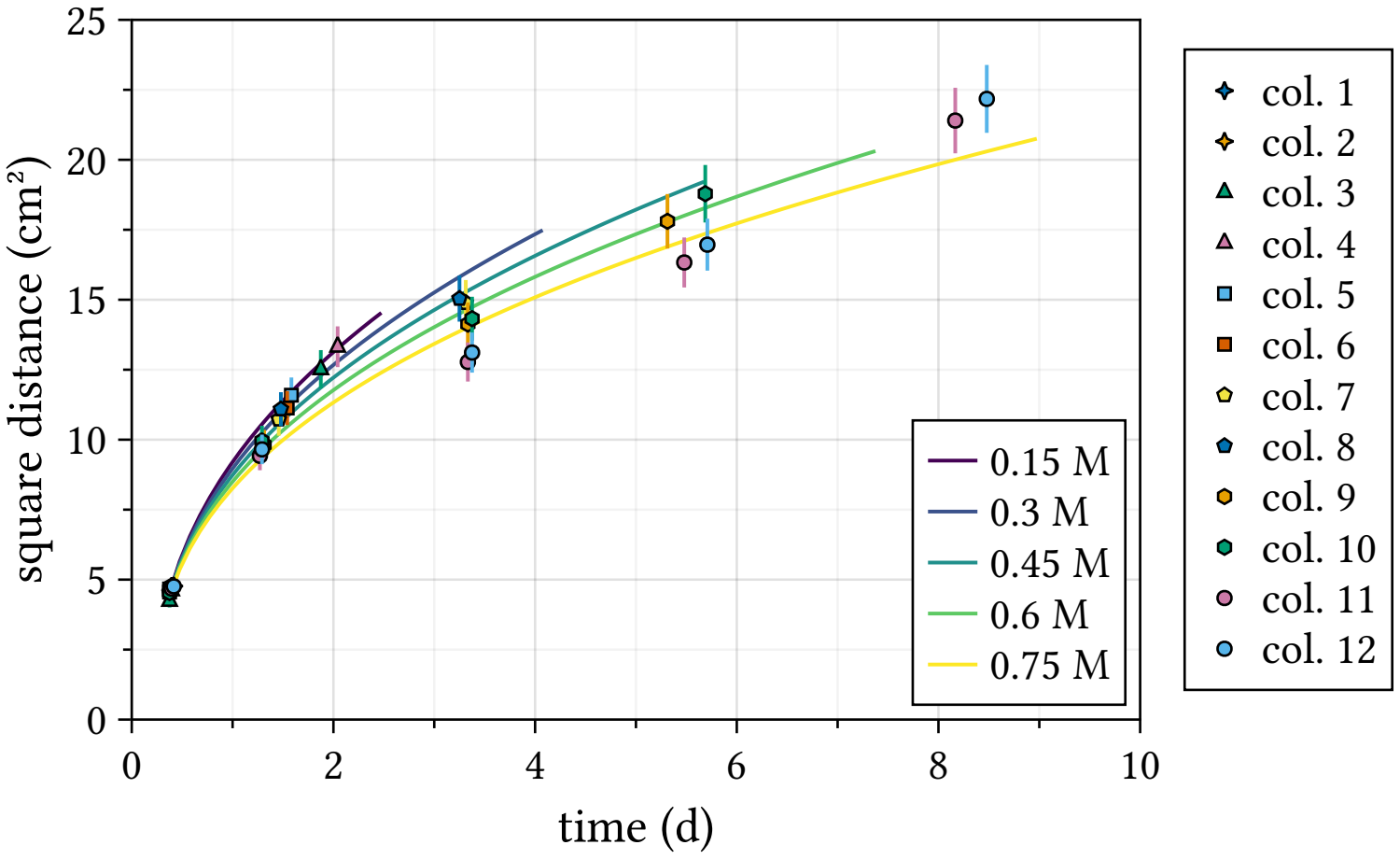

FIG. S5: **Titration of "top-added" buffer in ammonium chloride columns.** Medium composition: synthetic medium 2 (Tab. S2) + 18.7 mM $NH_4Cl$. **(a)** Kymographs. Top labels refer to the concentration of potassium phosphate pH 7 (K-phos7) in the 0.5 mL of solution added to the top interface. Figure 3c in the main text corresponds to columns 2, 8, and 12. **(b)** Scaling of band appearance coordinates. Different marker shapes characterize different amount of added buffer. Solid lines are the outcome of modeling the scaling using Eq. (S3) with [Glc] replaced by the buffer concentration, [K-phos7]. ML estimation of the parameters gives: $\alpha_0 = 0.44 \pm 0.02$, $\alpha_1 = -0.08 \pm 0.02$, $K_0 = 24.1 \pm 1.5\,\mathrm{cm^2 d^{-\alpha}}$, $\tau = 0.23 \pm 0.02$ d ($r^2 = 0.99$).

## V. MODEL OF PPF

### A. Response to parameter changes

We here report how observables associated with the pattern shape respond to control variables such as the concentration of glucose initially contained in the Glc-layer, $[\mathrm{Glc}]_0$, the concentration of buffer initially added at the top interface, $[\mathrm{B}]_0$, the cell proliferation rate, $k_\mathrm{p}$, and the rate of release of protons upon proliferation, $y_h$. In reference to Fig. S6, we report the response of the position of the bottom band and first top band, $x_b$ and $x_t1$ (1st and 2nd row), the average distance between bands, $\Delta x$ (3rd row), and the overall number of top bands, $N_t$ (4th row). The response of the pattern to increasing amounts of glucose (1st column), is qualitatively captured: all bands shift and are compressed upwards, with more bands appearing (Fig. 2c, main text). Increasing the concentration of the buffer solely affects the top bands by shifting and compressing them downwards. This is captured by our model, but not necessarily the fact that the number of bands increases, *cf.* Fig. S6c and 3c, the latter in the main text. However, we note that the number of top bands does not necessarily change homogeneously: With glutamine as the nitrogen source (Fig. 4b), different concentrations lead to non-monotonic changes of bands. Higher proliferation rates shift bands downwards (Fig. S6c and g). This is consistent with what can be inferred from experiments comparing glutamate and $NH_4Cl$ as nitrogen source, the latter being preferred by many bacteria (Fig. 2a vs 3c). The major effect of an increased rate of release of protons seems to be the increased number of bands (Fig. 4c, 2nd column). This is surprisingly consistent with our model (Fig. S6, 4th column.) Finally, our model is also qualitatively consistent with experimental observations of the total disruption of band formation in $NH_4Cl$ columns with no buffer added from the top interface (Fig. S7).

| parameter | symbol | value | notes |
|---|---|---|---|
| glucose diffusion coefficient | $D_g$ | 0.02 $cm^2$ /h | BN104089 |
| protons diffusion coefficient | $D_h$ | 0.3 $cm^2$ /h | BN106702 |
| base diffusion coefficient | $D_b$ | 0.03 $cm^2$ /h | based on hydrogen phosphate, $[HPO_4]^{2-}$ [24] |
| rate of transition to proliferation | $k_\mathrm{r}$ | 1 $h^{-1}$ | |
| rate of proliferation | $k_\mathrm{p}$ | 0.4 $h^{-1}$ | based on value established for fermenting *E. coli* [25] |
| rate of transition to quiescence | $k_\mathrm{q}$ | $10^2$ $h^{-1}$ | |
| rate of neutralization | $k_\mathrm{n}$ | $10^4$ $h^{-1}$ | |
| glucose consumption coefficient | $y_g$ | 4 g (Glc) /g (DW) | value found for *E. coli* [25]: 2–8 g (Glc) /g (DW) |
| protons production coefficient | $y_h$ | 0.005 g ($H^+$) /g (DW) | value found for *E. coli* [25]: 0.1 g ($H^+$) /g (DW) (*) |
| threshold of glucose | $g^*$ | $1.8 \cdot 10^{-8}$ g (Glc) /cm | based on value established for *E. coli* [26] |
| threshold of protons, transition to proliferation | $h^*$ | $2.6 \cdot 10^{-8}$ g ($H^+$) /cm | corresponding to approx. pH = 4.8 (**) |
| threshold of protons, proliferation | $h^{**}$ | $5 \cdot 10^{-8}$ g ($H^+$) /cm | corresponding to approx. pH = 4.5 |
| initial glucose concentration | $g_0$ | 0.03 g /cm | corresponding to approx. 2% (w/v) |
| initial proton concentration | $h_0$ | $2 \cdot 10^{-10}$ mol /cm | corresponding to approx. pH = 7 |
| initial base concentration | $b_0$ | $1.1 \cdot 10^{-3}$ mol /cm | corresponding to approx. 0.75 M of phosphate buffer |
| initial cell density | $c_0$ | $6 \cdot 10^{-7}$ g (DW) /cm | corresponding to approx. $3 \cdot 10^6$ cell /cm (***) |
| position of the bottom edge of the column | $x_\mathrm{E}$ | −2.2 cm | |
| position of the bottom interface | $x_\mathrm{B}$ | 0 cm | |
| position of the top interface | $x_\mathrm{T}$ | 6.95 cm | |
| height of the layer of top-added buffer | $\Delta x$ | 0.3 cm | corresponding to 500 µl of solution |

TABLE S6: **Parameter values**. Parameters found in volumetric densities are transformed to linear densities using the cross sectional surface of the experimental tubes, $A \simeq 1.5$ $cm^2$. (*) One may notice that the value of $y_h$ that we employ is substantially smaller than that of *E. coli*. This is consistent with the fact that *S. marcescens* are capable of 2,3-butanediol fermentation, a fermentation pathway that releases weaker acids compared to *E. coli*'s mixed acid fermentation [27]. (**) This value is set to a slightly smaller value than the lower bound of pH range of *S. marcescens* growth [28]. (***) DW stands for *dry weight*, and its value is based on $2 \cdot 10^{-13}$ g (DW) /cell, viz. within the range found for *E. coli* during stationary phase [29]. BN refers to BioNumber, see [30].

### B. Analysis of band formation

We finally broaden the scope of our model and ask what conditions discriminate the occurrence of band formation. However, answering this question presents some issues: the model is highly nonlinear and characterized by many parameters. On top of that, no unique operational definition of "band" exists.

To address these issues, we first simplify the model by disregarding the buffer diffusing from the top interface (Eq. B1c, main text), as well as the transition to quiescence (Eq. B2c, main text). Although these two processes are necessary for formation of certain bands (Fig. 5c, main text), they are not essential for band formation *per se* (Fig. 5b, main text). Regarding the definition of "band", we use an empirical definition inspired by our experiments and define it as a peak of the relative final-time log cell

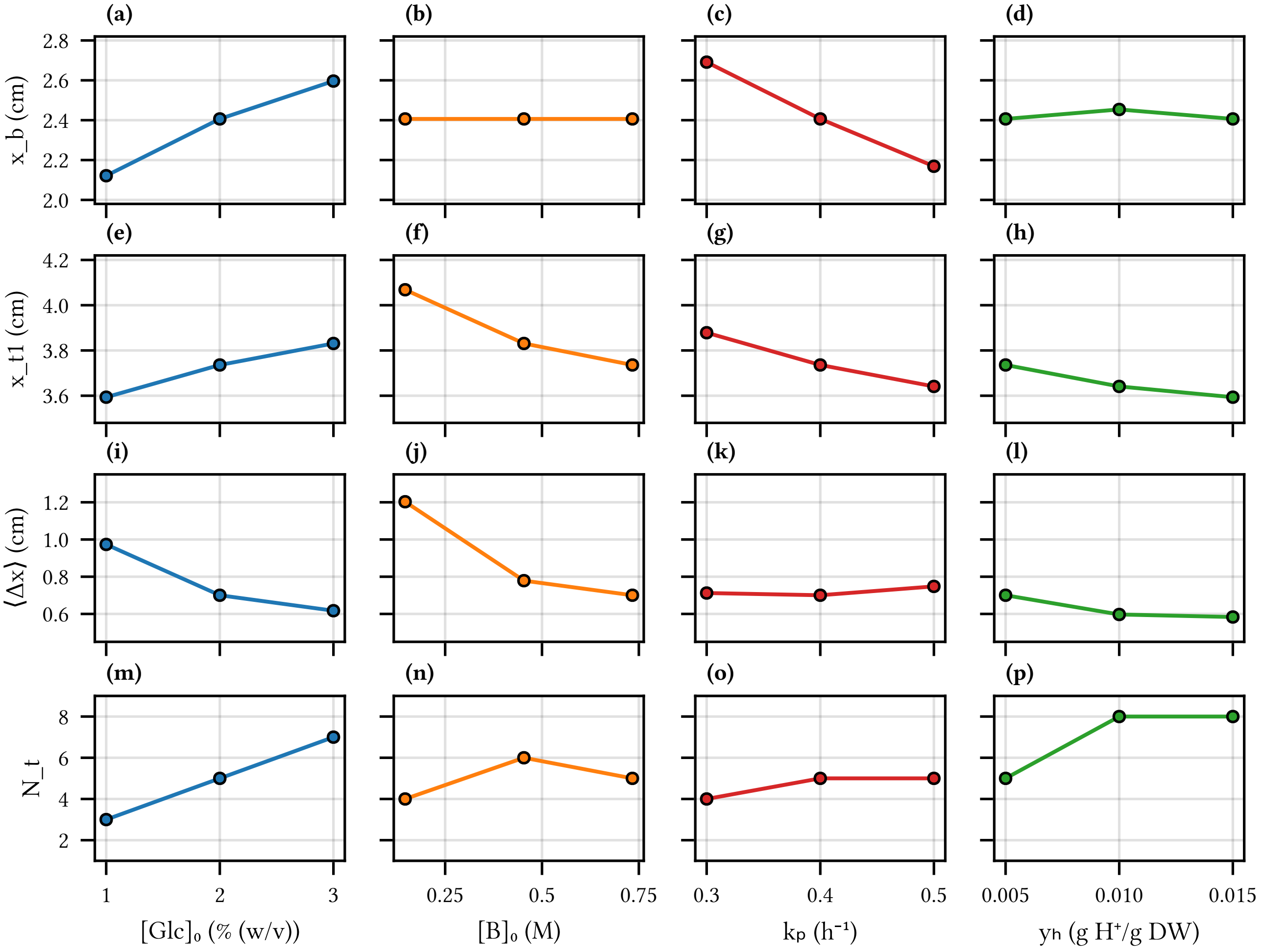

FIG. S6: **Response of the model to experimentally reproducible perturbations.**

density profile,

$$\ell(x) := \log \frac{s_0(x, t_f) + s_1(x, t_f)}{c_0}, \tag{S6}$$

higher than 1.5 and whose prominence is larger than a certain value, which we will set to either 0.5 or 0.75. Finally, it is conceptually convenient to lift the interpretation of $h$ as the concentration of protons, and interpret it as the concentration of a generic inhibitor molecule. Similarly, $g$ can be thought of as a generic activator.

In this simplified and defined setting, we introduce two variables that statistically discriminate band-forming systems from band-free ones. This result is presented in the phase diagrams in Fig. S8, where these variables vary along the axes. The horizontal axis represents the ratio of inhibition thresholds, $h^*/h^{**}$. The smaller this value is, the larger is the window of inhibitor concentrations that enables cellular proliferation but not the transition to proliferation. The vertical axis can be interpreted as the ratio of two diffusion length scales:

$$\Lambda := \sqrt{\frac{D_h}{k_p\, y_h}} \left[ \sqrt{\frac{D_g}{k_p}} \right]^{-1} = \sqrt{\frac{D_h}{D_g\, y_h}}, \tag{S7}$$

the typical distance traveled by the inhibitor within the time scale in which it is produced, over the typical distance traveled by the activator within the time scale of proliferation. Loosely speaking, this variable is a (dimensionless) *scale of lateral inhibition*: larger values correspond to longer ranges in which the inhibitor can operate compared to the activator.

Each point in Fig. S8 is the result of a simulation with key parameters chosen randomly within biologically-plausible ranges (details in caption). Orange (resp. blue) points represent simulations in which at least one (resp. no) band appears. The shaded areas in the background discriminate regions in which bands typically appear (orange) or not (blue, details in caption). The difference between Fig. S8a and Fig. S8b lies in the choice of minimal prominence defining a band (0.5 and 0.75, respectively).

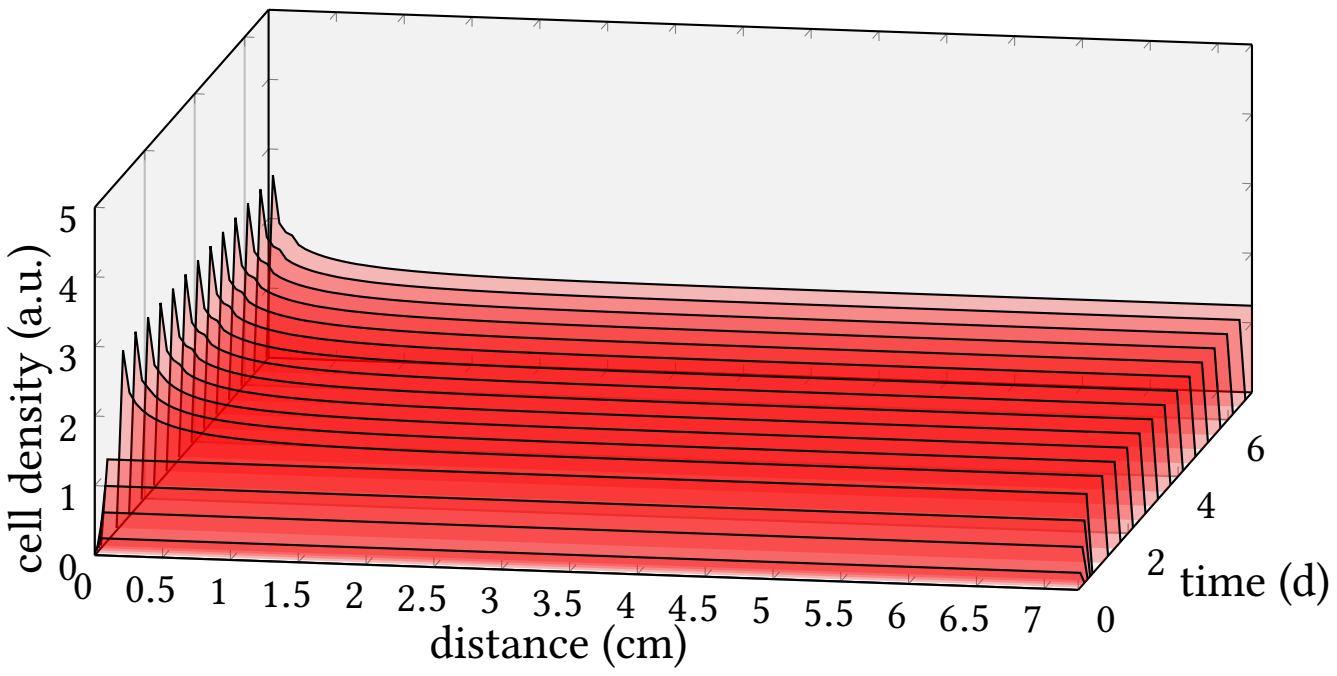

FIG. S7: **Model simulation with** 0.01 % **(w/v) glucose initially added to the C-layer and no buffer from the top interface**. The concentration of glucose initially present in the C-layer is set to $1.5 \cdot 10^{-4}$ g /cm, which corresponds to 0.01 % (w/v). All the other parameters are set as in Tab. S6.

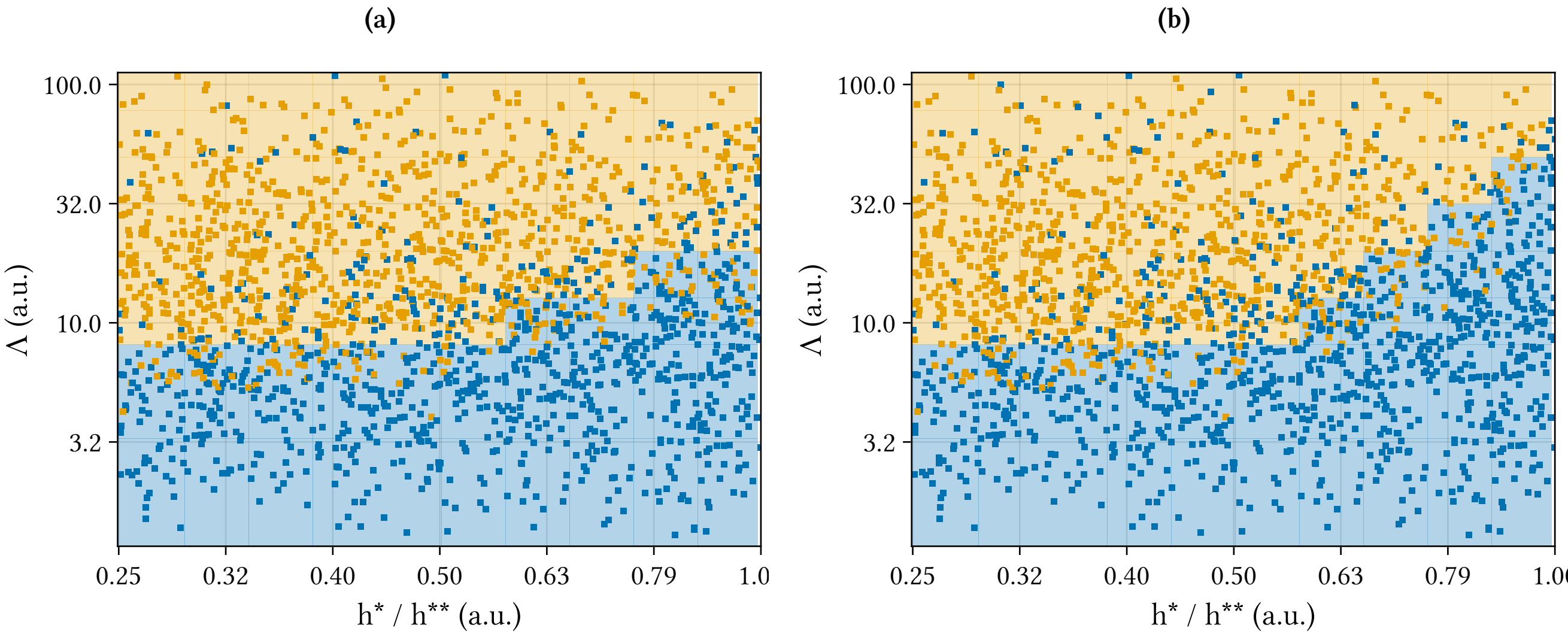

FIG. S8: **Band formation phase diagram**. Each point (1963 in total) corresponds to one system in which the following parameters are randomly sampled within biologically-plausible ranges. Depending on whether a particular parameter's range would span one or more order of magnitudes, we sampled from a uniform or log-uniform distribution, respectively. We made an exception for $h^*$, since we care more about its logarithm (e.g. just as pH is a logarithm). The distributions and ranges used are, where the units are as in Tab. S6: $k_\mathrm{p} \sim$ Uniform$(0.1, 0.7)$; $y_g \sim$ Uniform$(1, 10)$; $y_h \sim$ LogUniform$(0.01, 1)$; $k_\mathrm{r} \sim$ LogUniform$(0.01, 100)$; $D_h \sim$ LogUniform$(0.003, 0.3)$; $h^* \sim$ LogUniform$(1.26 \cdot 10^{-8}, 5 \cdot 10^{-8})$; $g^* \sim$ LogUniform$(1.8 \cdot 10^{-10}, 1.8 \cdot 10^{-6})$; and all the other parameters are as in Tab. S6. Band-forming systems are colored orange, whereas band-free ones are blue. The ratio $h^*/h^{**}$ varies along the horizontal axis, whereas $\Lambda$ (Eq. (S7)) along the vertical axis. For convenience, the $\Lambda$-axis is rescaled by a factor $\sqrt{h_0/c_0}$, and the scale of both axes is logarithmic. At such scale, the phase space is divided in $\sim 0.060 \times 0.20$ sized blocks ($10 \times 10$ total). For each block, we assess whether the majority of systems are band-forming or band-free and mark the background accordingly (orange vs blue, respectively). In **(a)**, the minimal prominence of cell density peaks (Eq. (S6)) that define bands is set to 0.5, which is one third of the minimal peak height (1.5), whereas in **(b)** the minimal prominence is 0.75.

The insight gained from Fig. S8 is the following: within the range of parameters sampled, band formation typically occurs when the scale of lateral inhibition ($\Lambda$) is substantially large. The boundary separating this regime from the band-free one depends on the ratio of inhibition thresholds, but it does so solely for small ratios ($h^*/h^{**} \lesssim 0.6$). Also, imposing more stringent conditions on the definition of band (such as in Fig. S8b) does not alter the boundary value for small ratios.

We conclude with three remarks.

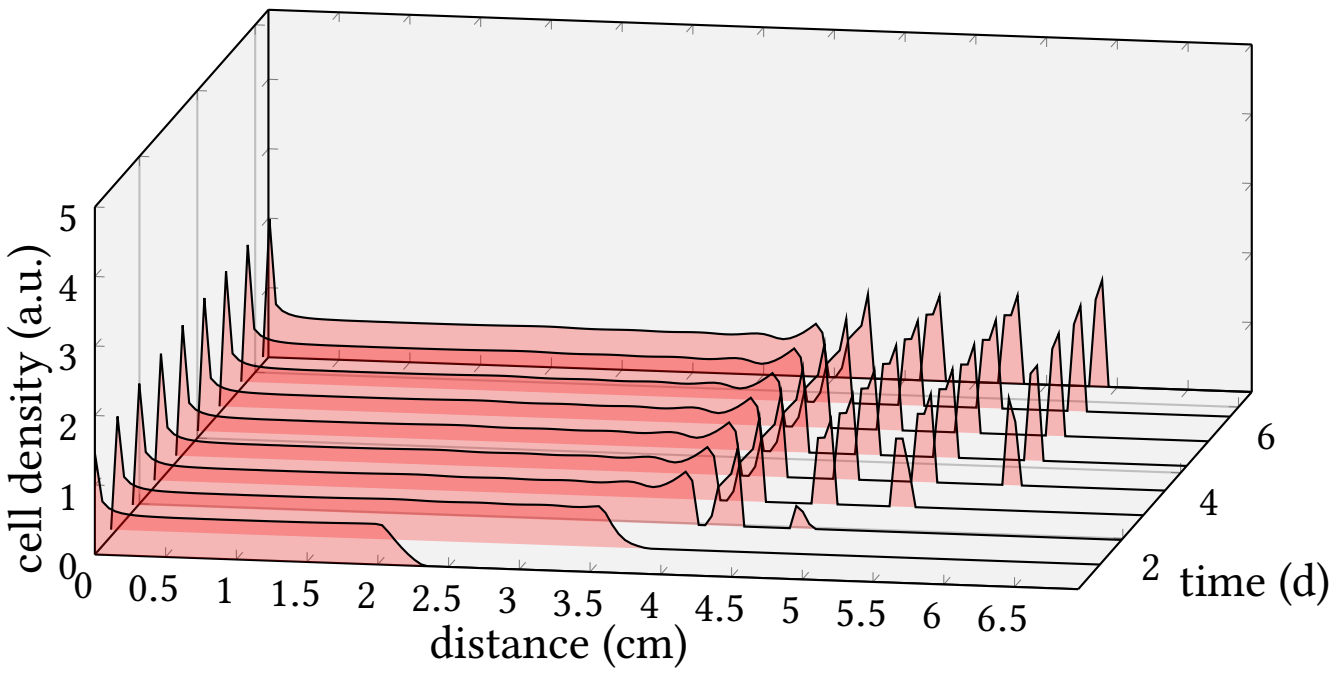

FIG. S9: **Simulation of the simplified model with balanced diffusivities**. This simulation is based on the simplified model described in §V B with parameters (given with 4 significant digits; units in Tab. S6): $D_h = 0.02036$, $k_\mathrm{p} = 0.6948$, $k_\mathrm{r} = 0.7195$, $y_g = 7.088$, $y_h = 0.03300$, $g^* = 2.269 \cdot 10^{-6}$, $h^* = 2.296 \cdot 10^{-8}$, and all other parameters as in Tab. S6.

First, we identified $\Lambda$ and $h^*/h^{**}$ as variables discriminating band formation, building on our intuition of the model. It is clear that variables better at discriminating band formation could be obtained through machine learning techniques. Whether or not such variables would have a clear interpretation and bring novel insights is an open question that we leave for a future study.

Second, $\Lambda$ discriminates the conditions for band-formation better than $\sqrt{D_h/D_g}$ would. To show this, one could compare two logistic regressions of band formation (response variable) using either $\Lambda$ or $\sqrt{D_h/D_g}$ as regressors, respectively. For the data plotted in Fig. S8, we find a smaller BIC for the former model (ratio of BIC values $\sim 0.9$), which suggests that $\Lambda$ is a better variable.

Third, bands can form even when the diffusivities of the activator and inhibitor morphogens have comparable values, $D_h \simeq D_g$ (Fig. S9), a condition which would not enable the formation of Turing patterns [31].

## ACKNOWLEDGMENTS

We thank the Julia Programming Language community for their work on the eponymous language, the Rockefeller University Genomics Resource Center for technical advice, use of equipment, and sequencing, and the Rockefeller University Precision Instrumentation Technologies Resource Center for technical advice and use of their laser cutter.

## VI. SUPPLEMENTARY MOVIES AND FIGURES

MOV. S1: **Time-lapse movie of *S. marcescens* ATCC 13880 growing in 12-replicate Phytagel columns**. Medium composition: synthetic medium 1 (Tab. S2) + 6.8 mM glutamate. The snapshots in Fig. 1 (main text), as well as the kymographs depicted in Fig. 2(a) (main text) and Fig. S1, are derived from this experiment. The white flame-like shapes near the bottom of the tubes are reflections of the light source on the glass.

MOV. S2: **Titration of glutamine and qualitative pH dynamics in Phytagel columns**. Medium composition: synthetic medium 2 (Tab. S2) + varying amount of glutamine. The last 6 columns are replicas of the first 6 but feature the addition of a pH dye (5 μg/ml chlorophenol red), which is violet at pH 6.7 and above and yellow at pH 4.8 and below. A white background is added to the right half of the tube rack to facilitate visualization of the pH dye color. The concentration of glutamine in the 6 pairs (± dye) of columns is 2.5, 5, 10, 15, 20, and 25 mM. The kymographs in Fig. S13 are obtained from the first 6 columns of this movie. Note how pH drops earlier close to the bottom interface, as well as the fact that a high pH persists close to the top interface, where cells catabolize amino acids (glutamine in this particular experiment), and release ammonia (a weak base).

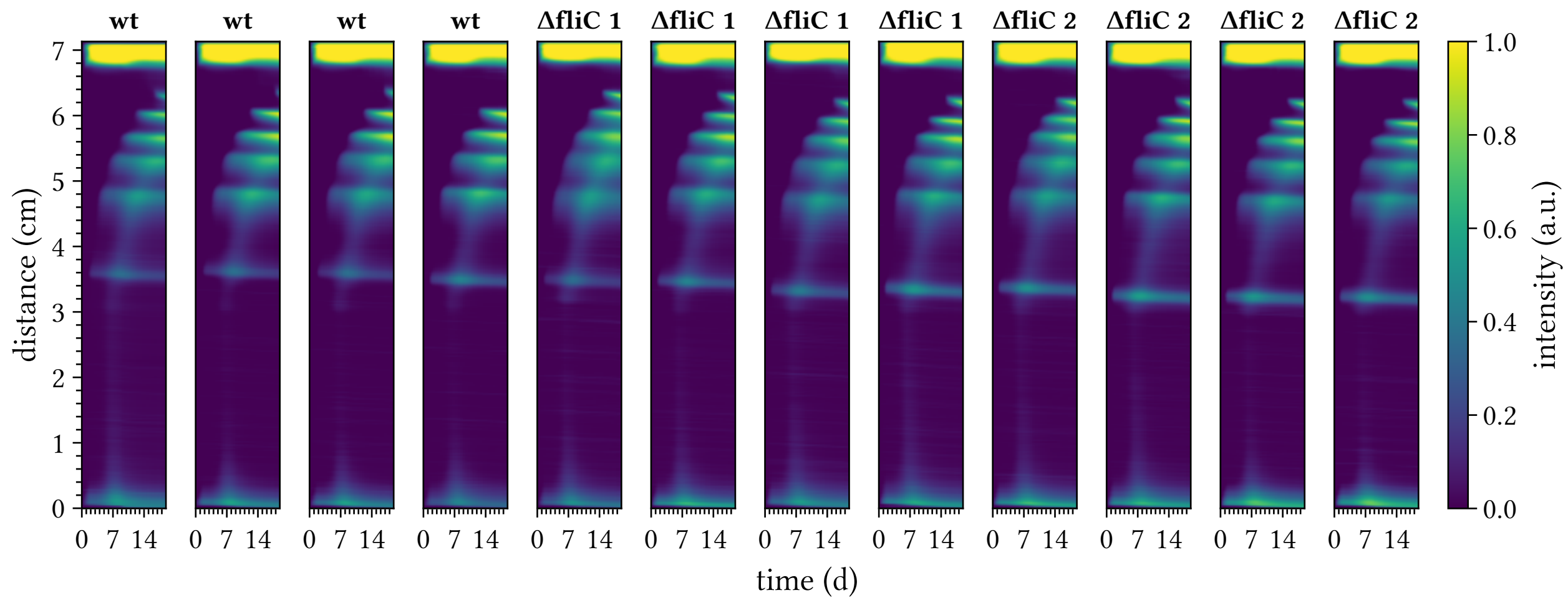

FIG. S10: **PPF occurs in non-motile flagellin mutants**. Medium composition: synthetic medium 1 (Tab. S2) + 6.8 mM glutamate. "wt" refers to the wild type *S. marcescens* (ATCC 13880) whereas "ΔfliC 1" and "ΔfliC 2" to two different isolates from the construction of isogenic ΔfliC mutants. These mutants lack flagellin, the major structural assembly unit of flagella and are hence non-motile, as verified in tests on motility agar. No difference in the band pattern is appreciable.

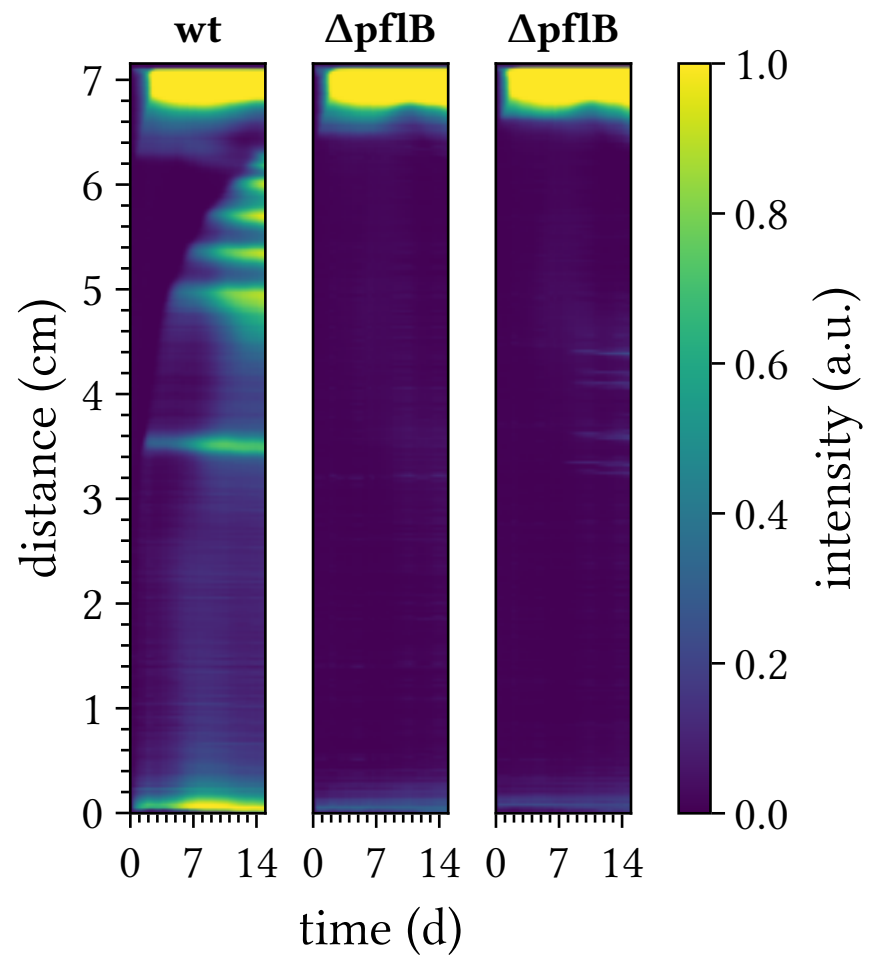

FIG. S11: **PFF is absent in a mutant affecting fermentation**. Medium composition: synthetic medium 2 (Tab. S2) + 6.8 mM glutamate. "wt" refers to the wild type *S. marcescens* (ATCC 13880), whereas "ΔpflB" refers to a mutant in pyruvate formate-lyase. This mutant is incapable of completing fermentation pathways, and hence incapable of growing anaerobically on glucose (although it can grow aerobically).

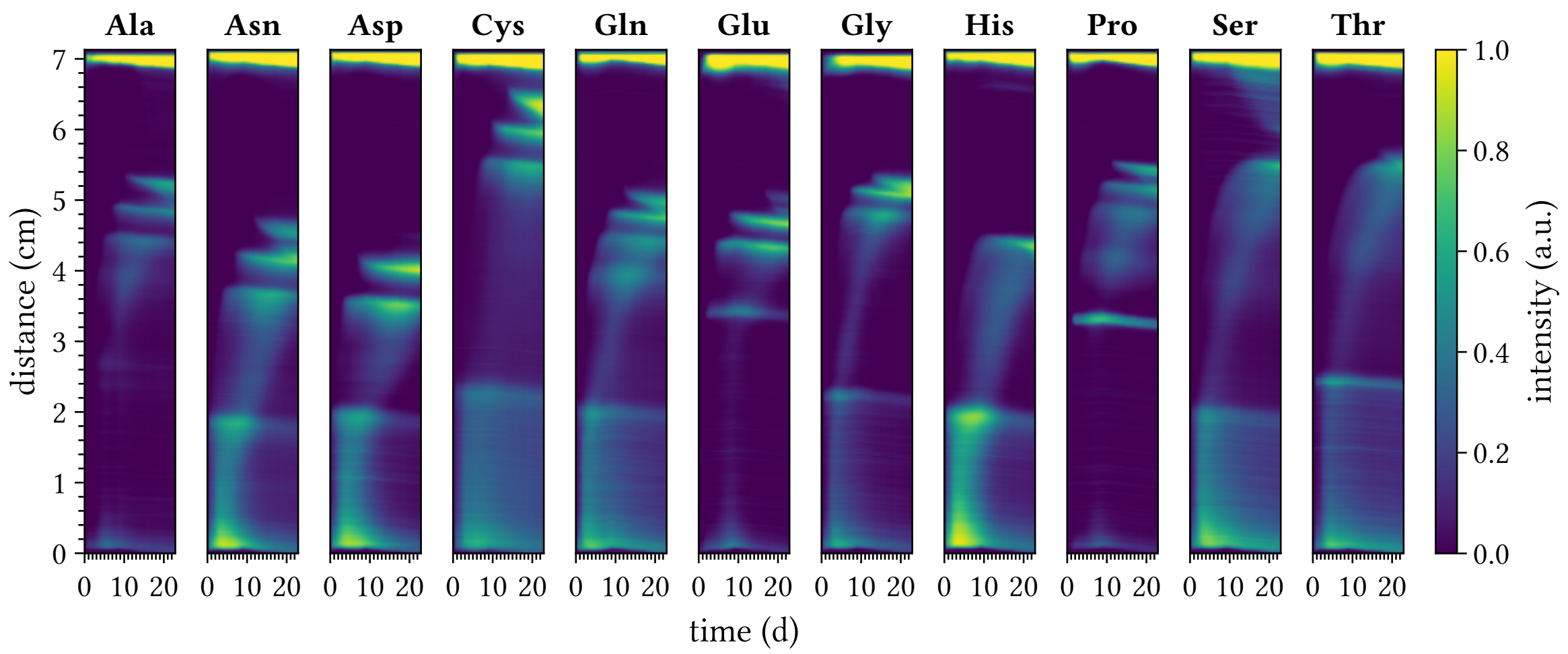

FIG. S12: **Single amino acid [*amino acids at an exhibition*]**. Medium composition: synthetic medium 1 (Tab. S2) + 10 mM of single amino acid (see top of kymographs). Ala, alanine; Asn, asparagine; Asp, aspartate; Cys, cysteine; Gln, glutamine; Glu, glutamate; Gly, glycine; His, histidine; Pro, proline; Ser, serine; and Thr, threonine.

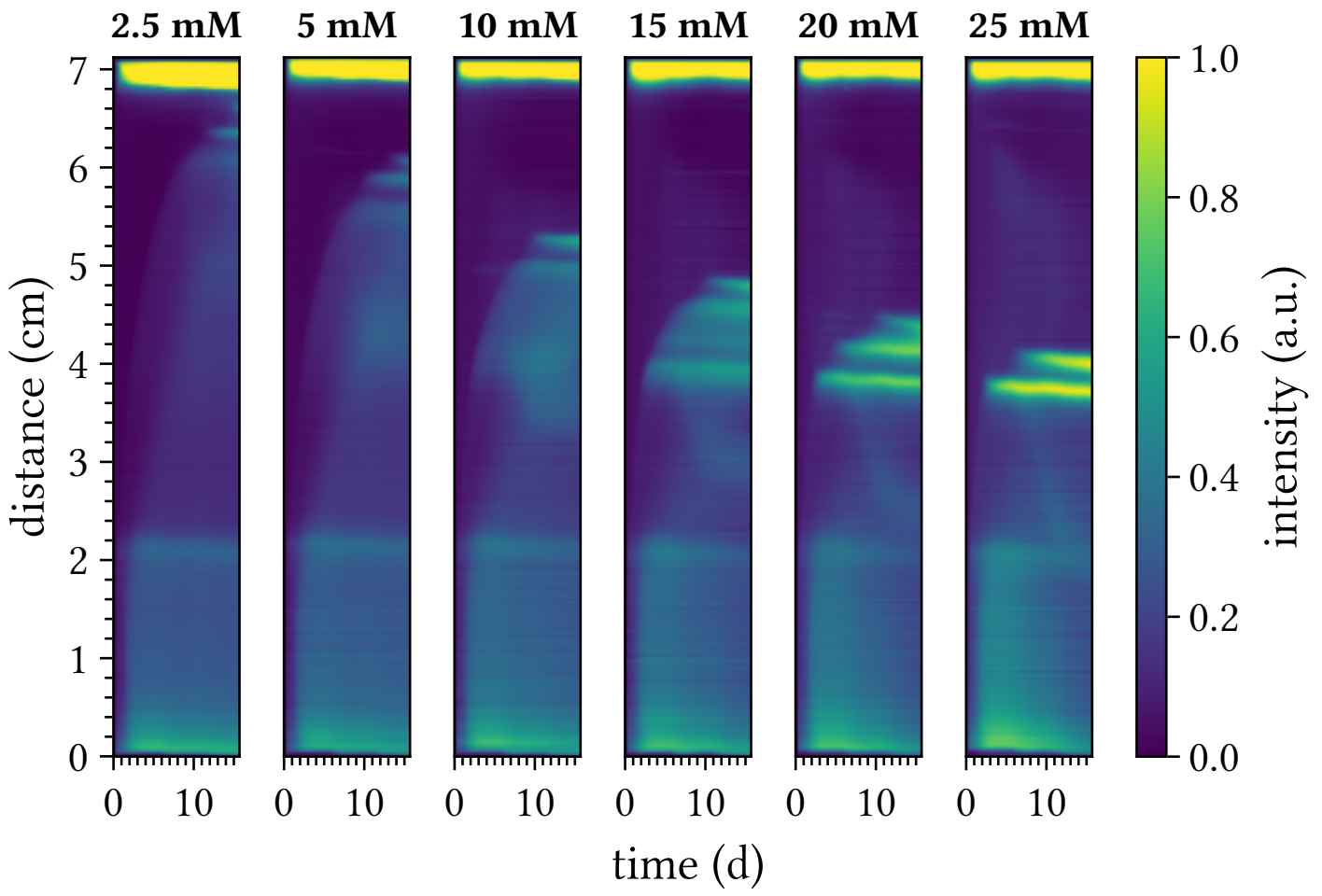

FIG. S13: **Titration of glutamine**. Medium composition: synthetic medium 2 (Tab. S2) + varying amount of glutamine (see top of kymographs). Note how the position of the bottom band is pretty robust against changes of glutamine concentrations. In contrast, the number of top bands and their position change dramatically.

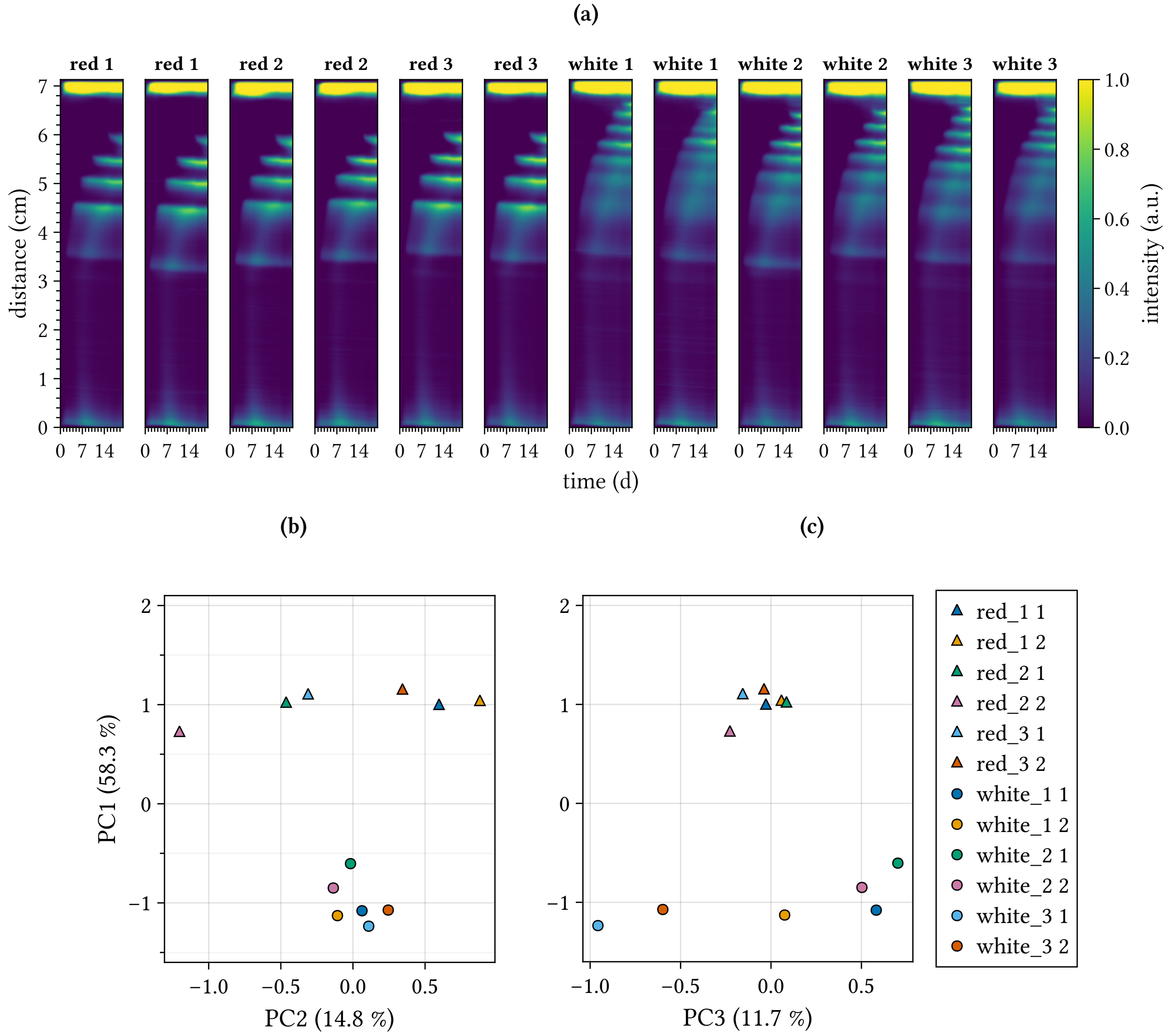

FIG. S14: **PPF originating from different *S. marcescens* colony phenotypes**. Medium composition: synthetic medium 1 (Tab. S2) + 6.8 mM glutamate. **(a)** Kymographs of columns inoculated with either of two phenotypes sampled from the same clonal population: Columns labeled with "red" and "white" are inoculated with cells sampled from red or white colonies, which reflects the colony's production of the red pigment prodigiosin. The different numbers label inocula from cultures prepared from different colonies. **(b)** and **(c)** Scatter plots of the first three principal components of the intensity profiles at the final time (proportion of variance explained: 84.8%). The first principal component (58.3%) clearly discriminates patterns generated by different phenotypes.

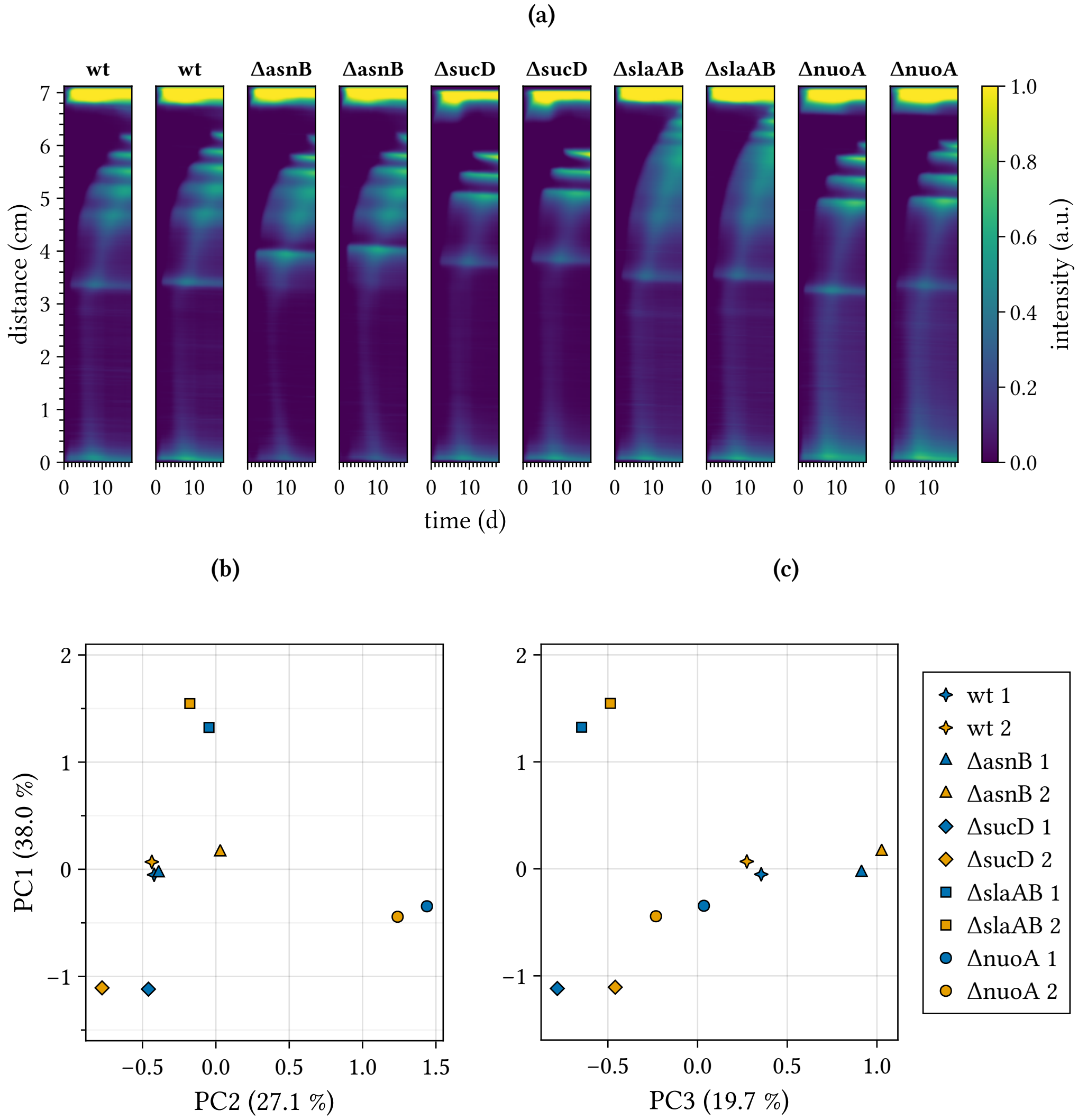

FIG. S15: **Knockout mutants**. Medium composition: synthetic medium 1 (Tab. S2) + 6.8 mM glutamate. **(a)** Kymographs of PPF generated by different knockout mutants. "wt" refers to wild type. *ΔasnB* mutants cannot synthesize the protein asparagine synthetase B, which is involved in several amino acid metabolic pathways [32]. *ΔsucD* mutants cannot properly synthesize succinyl-CoA synthetase, an enzyme involved in the TCA cycle [33]. *ΔslaAB* mutants cannot properly synthesize enzymes involved in the 2,3-butanediol fermentation pathway (2-acetolactate synthase, and 2-acetolactate decarboxylase) [27]. *ΔnuoA* mutants cannot properly synthesize NADH:quinone oxidoreductase, which is involved in respiratory metabolism [34]. **(b)** and **(c)** Scatter plots of the first three principal components of the intensity profiles at the final time (proportion of variance explained: 84.7%). Although certain pairs of pattern appear similar (such as those generated by the wild type strain and the *asnB* knockout mutant), principal components discriminate the patterns generated by different mutants.

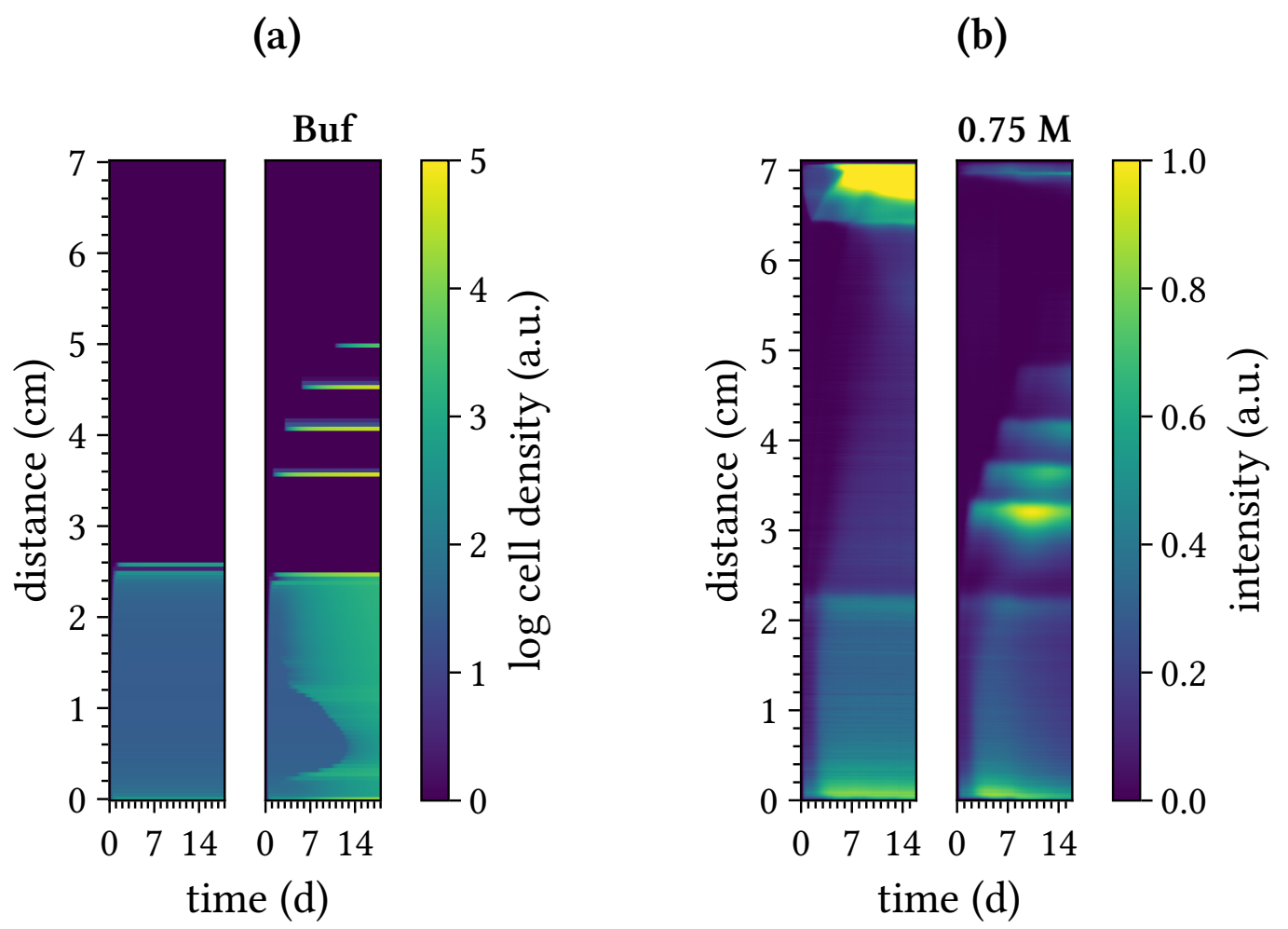

FIG. S16: **Model results vs experiments. (a)** Results of simulations presented in the form of kymographs rather than ridge plots (*cf.* Figs. 5b and 5c in main text). **(b)** Replica of two kymographs in presented in Figs. 3c, which correspond to the columns whose conditions are simulated in (a).